\documentclass{article}
\usepackage[letterpaper, left=1in, top=1in, right=1in, bottom=1in,nohead,includefoot]{geometry}
\usepackage[svgnames,x11names]{xcolor}\def\b{\color{RoyalBlue4}}\def\k{\color{black}}  
\usepackage{graphicx,placeins,charter,natbib,bibentry,etaremune,caption}  
\RequirePackage[colorlinks,citecolor=blue,urlcolor=blue]{hyperref}
 
\begin{document}
 
\begin{center} 

{\bf\Large A Conversation with Mike West}

\bigskip

{\large  Hedibert F. Lopes (INSPER S\~ao Paulo) and Filippo Ascolani (Duke University)}

\bigskip

Updated January 2026 (Original August 2025)

\end{center}

\begin{abstract}
Mike West is currently the Arts \& Sciences Distinguished Professor Emeritus of Statistics \& Decision Sciences at Duke University.  Mike's research in Bayesian analysis spans multiple interlinked areas: theory and methods of dynamic models in time series analysis, foundations of inference and decision analysis, multivariate and latent structure analysis, stochastic computation and optimisation,  among others. Interdisciplinary R\&D has ranged across applications in commercial forecasting, 
dynamic networks, finance, econometrics, signal processing, climatology, systems biology, genomics and neuroscience, 
among other areas. Among Mike's currently active research areas  are forecasting, causal prediction and decision analysis in business, economic policy and finance, as well as in personal decision making.

Mike led the development of academic statistics at Duke University from 1990-2002, and has been broadly engaged in professional leadership elsewhere.  He is past president of the International Society for Bayesian Analysis (ISBA),  
and has served in founding roles and as board member for several professional societies, national and international centres and institutes.  Among awards for research and professional service,   Mike is a three-time winner of the Mitchell Prize for applied Bayesian statistics research:  in 1994, for statistical 
modelling and analysis in pal{\ae}oclimatology; in 1997, for statistical modelling in neurophysiology; and in 2012 with co-authors, for statistical modelling in spatio-dynamic systems immunology.   In 2014, Mike was an inaugural recipient of the ISBA Zellner Medal, recognising \lq\lq{\em 
a scientific life spent always at the top, and a vision of the future which became reality}''; in 2018, he was recognized with the Akaike Memorial Award as a \lq\lq {\em great pioneer in the field of Bayesian statistics}''; in 2022, he received the Bruno de Finetti Award as \lq\lq{\em an 
outstanding scholar who has provided significant contributions to the advancement of Bayesian Statistics}''; and in 2025 he received the NISS  Sacks Award that \lq\lq{\em recognizes sustained, high-quality cross-disciplinary research involving the statistical sciences}'' and Mike's \lq\lq{\em outstanding contributions to Bayesian statistics, time series analysis, and computational methods across various fields in applied science, including economics, environmental sciences, and biomedical research}''. 

Mike has been active in research with various companies, banks, government agencies and academic centres, 
co-founder of a successful biotechnology company, and board member 
for several  financial and  IT companies. He has published 4 books, several edited volumes and  over 200 papers.
Mike has worked with many undergraduate and Master's research students, and as of 2025 has mentored    
around 65 primary PhD students and postdoctoral associates  who moved to academic, industrial or governmental positions involving advanced statistical and data science research. 

\end{abstract}

\section{Early years}

\textbf{Hedibert and Filippo}: Mike, would you tell us a bit about your childhood and what interests you had growing up?

\textbf{Mike}: I was born in the North East of England, in what is nowadays the pleasant coastal enclave of Tynemouth.   I spent my childhood in several places in UK;  my father was in the British Army and we moved around ... a lot. I guess I benefited from a diversity of  cultural exposures as a result, in spite of the inevitable fragmentation of community and educational trajectories due to frequent relocations.  I was always engaged in art at home, and was fortunate that several of the places we lived were in fairly rural areas so I enjoyed spending much time in the natural environment.  At school, my main interests were art, history and maths, then later languages.  I enjoyed maths uniformly throughout my  school years. Then, while I have always been interested in the natural world and natural sciences,   I never developed much of an academic interest in building \lq\lq stores of knowledge'' (a.k.a. memorizing facts);  biology and chemistry books are big (and expensive) while maths and arts are all about, well, just building on a few basic \lq\lq principles'' and-- primarily-- making it up at the time, as needed, and as inspirations arise!   

\textbf{Hedibert and Filippo}: When did you encounter mathematics and statistics for the first time? 
 Did you have anyone of inspiration towards the academic career? E.g. a relative, a teacher,..., and why did you decide to enrol in a mathematics degree?
 
\textbf{Mike}: I don't recall any particular \lq\lq influencers'' academically through my middle and high-school years,  though I did have a couple of really fun maths and art teachers.   My family and the communities I grew up in were remote from academia, with limited historical recognition of university as a potential post-school path. My parents, however,  did promote educational advancement and,  at some point, it became clear to me that maths was a way ahead. University was then (as it has been for many) a path of least resistance, to a certain extent. 
I did not encounter statistics, or even probability, at school.  I had an initial brief taste of introductory statistics at university, but quickly decided it was not for me!  Group theory and quantum theory were just more, well, new and different.    Then, during the summer before my 3rd and final undergraduate year, I landed a summer job in a bank. Not an investment bank that so many of our (and my advisee) students seek and find nowadays, but a national high street bank.  I had very little to do there. Conscious that I might be interested in jobs in other areas in just a year or so, I spent a lot of time that summer just reading, including a few statistics books. 
I have to say that I did not find reading frequentist statistics texts easy.    Then, at the start of my final undergraduate year I met the new statistics professor, Adrian Smith; his take on statistics was, well, a little different, and opened up new reading.  The maths was the kind I liked; but it was the mix of philosophy and the politico-scientific context of Bayesian statistics in the late 1970s that became a little more engaging. I took a final-year undergraduate course with Adrian and that set me up to consider doing a bit more, resulting in my decision to take up Adrian's kind offer to  follow-on with a PhD at the University of Nottingham.  I will add that I also took a final-year undergraduate course in time series with Paul Newbold, and that partly influenced me in thinking with Adrian about PhD research areas.  These were  my only two undergraduate statistics experiences that positively mattered, and were clearly most influential. Then, of course, almost all of what I really know and have taught about in statistics over the decades since really came about by learning-through-doing! 

\textbf{Hedibert and Filippo}: You did your PhD with Adrian Smith. How were your interactions?

\textbf{Mike}: Typically over a beer in a local pub.  In those days, PhD in UK mathematics departments was rather unstructured; no courses, just jump-in and figure out a way ahead. A lot of library time meandering through statistics and signal processing journals and literature, and of course  trying to figure out the bigger picture of statistics since I had very limited formal exposure to that point.   I had a good deal of frustration with the (in)ability to compute. One developing interest was to do integrations for Bayesian analysis  in sequential time series settings. The then-current numerical analysis methods were focused on high precision in one dimension; we, of course, wanted any decent precision at all in at least a few dimensions.  So creative approximations were {\em de rigueur}.    Adrian was then very involved in thinking about computational challenges more broadly, and through him I had useful interactions with others engaged in developing advances for Bayesian computation generally-- preceding by a decade or so the simulation-based revolution.  Of course, 
interacting with Adrian also led to broader connections to the professional community, including with Dennis Lindley who later became an important influence on aspects of my research too.

My own interests and PhD research moved in other directions, more exclusively in time series modelling and methodology 
for forecasting and monitoring.  It was here that my interactions with Adrian, and his roles in my early steps into serious statistical R\&D, were   focused and productive. I joined Adrian's collaboration with biomedical scientists in studies of post-operative monitoring of renal transplant patients; this was   my first serious involvement in real-world statistics, and had a main impact in my PhD development on both methodological and applied sides.  It was a very productive collaboration in terms of core statistical R\&D as well as implementation in the applied setting, and a first context for me in collaborating with non-statistical professionals in their domain. The opportunity to cut ones professional teeth in a detailed interdisciplinary collaboration as a PhD student was not so common in those days, while it is so key and critical in terms of early experiences and, of course(?), front-and-centre in most PhD environments nowadays.

\textbf{Hedibert and Filippo}: Your research in the first years after PhD alternated between applications and some papers more focused on methodology and computation. Who and what were your main sources of inspirations at the beginning?

\textbf{Mike}: At the start of the 1980s, main research influences for new researchers in UK were largely \lq\lq local'' colleagues, visitors/conferences and the published literature. Email was only just beginning (as a within-institution communication mode) and the internet was years away.  Several months before I completed my PhD, I was fortunate to move to the faculty at Warwick University, a very small (5!) but top statistics department with a strong Bayesian outlook. Routine interactions with senior Bayesians Jeff Harrison and Tony O'Hagan naturally had substantial impact in those early years. 
 Jeff Harrison was a very main influence, and of course became a long-term collaborator and close personal and family friend. In addition to my own expanding methodological interests, with and through Jeff I became involved in one or two collaborative projects in industry, implementing and extending Bayesian methods in commercial forecasting and monitoring applications. I branched out with additional commercial consulting-style R\&D with other industry groups (multinational revenue forecasting, market research, etc.), and, frankly, that's where I really started to learn statistics.  Much of my resulting academic research in theory and methods gained direction and perspective from that.   An important period was our so-called Bayesian Study Year in 1985-6. This was conceived and initiated by Tony O'Hagan and  run as a program with short- and long-term visitors from academia and industry over several months.  I was designated an industry lead for fund raising, which helped me push out to make industry connections.  That year generated a range of connections and research inspirations, with a number of longer-term visiting academic statisticians-- some eventual life-long friends and several that were influential in my broadening views and interests in statistics as well as my career path thereafter. With a number I could add, some of the more engaging visitors that year-- as well as senior researchers I met and interacted with at conferences in the early 1980s-- included Jim Berger (then at Purdue), Morrie DeGroot (Carnegie Mellon), Art Dempster (Harvard), Bill DuMouchel (then at Bell Labs.), Bruce Hill (Minnesota), Wolfgang Polasek (then at Basel), Mark Schervish (Carnegie Mellon) and Nozer Singpurwalla (George Washington), as well, of course, as Dennis Lindley.

At its best, statistics thrives on the fertility of the interplay between applications and core theory and methods; advancing and customising statistical models and methods for specific applications feeds back to stimulate conceptual and theoretical developments of broader interest. This developed for me  in those early years, and has remained a cornerstone of my professional philosophy. I will add that teaching was then, and remained, a major contributor to learning and exploring new areas related to-- or potentially related to-- research. This develops through the continual need to \lq\lq better understand'' and \lq\lq learn things (a)new'' in teaching preparation, and centrally through the rethinking motivated (and often forced!) by students-- at all levels-- questioning you and calling out when you \lq\lq screw-up'' in class. I learned a great deal related or relevant to my research though teaching in those early 7 years at Warwick, and continued to do so through 36 years teaching at Duke.

\begin{figure}[htbp!]
\centering 
\begin{minipage}{3.2in} 
\centering
\includegraphics[clip,width=3.0in]{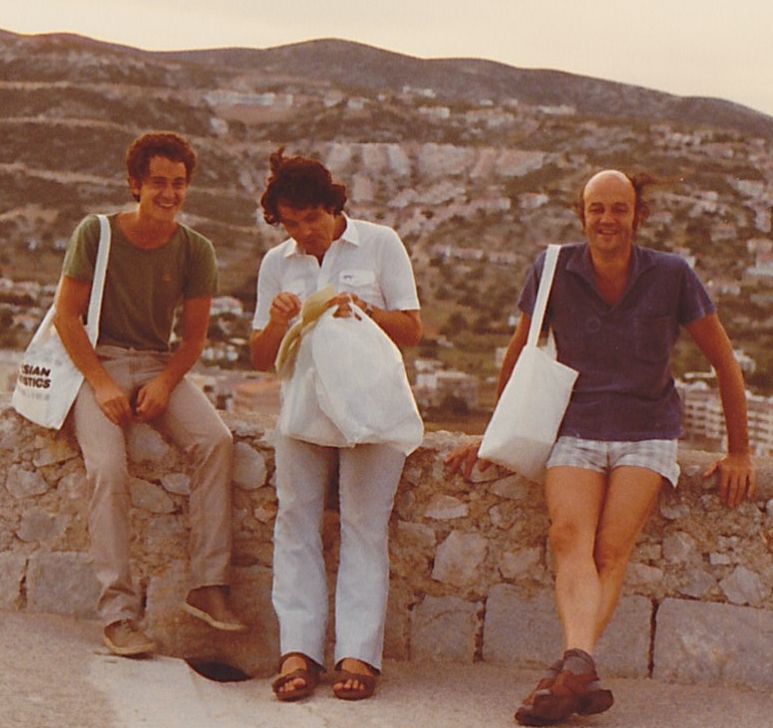}  
\caption*{Mike,  Jim Smith and Jeff Harrison at the 2nd Valencia International Meeting in Spain, 1983.
}
\end{minipage}
\hskip.25in
\begin{minipage}{3.2in} 
\centering
\includegraphics[clip,width=1.5in]{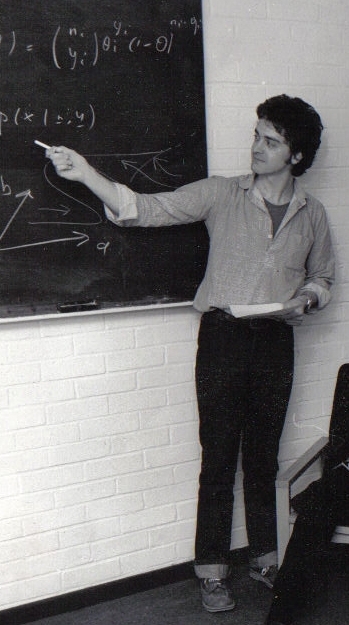}  
\caption*{Mike holding forth, in the Department of Statistics, University of Warwick, 1982.
}
\end{minipage}
\end{figure}

\section{ISDS and Duke University}

\textbf{Hedibert and Filippo}: You joined Duke in 1988, six years after your PhD. How did this happen?  

\textbf{Mike}: As a new researcher I met and developed interactions with a number of senior Bayesians at and following a couple of important conferences: the 1982
 Institute of Statisticians Meeting on Practical Bayesian Statistics in Cambridge in 1982, and the Valencia International Meeting in 1983.  
Some of these interactions generated follow-on invitations to visit, including a number of   groups in USA.   Coupled with these developments, in early 1984 I received a phone call from Carl Morris, then one of the editors of JASA, inviting Jeff Harrison, Helio Migon and I to the August Joint Statistical Meetings to present our just-accepted paper on dynamic generalised linear models as the JASA Theory~\& Methods invited discussion paper that year. Art Dempster was one of the invited discussants, and so things came together to do a little US tour through several Bayesian groups in US in 1984-- including Harvard statistics and the Berkeley industrial engineering/OR group (an important Bayesian group with Dick Barlow, Bill Jewell, Bob Oliver and others).  That began a number of connections that grew through the Bayesian Study Year (noted above) and evolved into conversations about repeat, perhaps longer \lq\lq visits''.   This was in the early 1980s economic realities in UK and US, and in particular at a time when UK academia had been, and continued to be, under political and economic stresses that-- while perhaps not so extreme as in more recent decades-- were, well, constraining. 

Following additional outreach from various statistics groups in US, and visits to top Bayesian centres in 1986/7,  Duke emerged as ... well, interesting. Duke had no statistics department!  And a less than robust history of engagement with statistics as anything other than a perceived-to-be necessary (though not always effected)  ingredient for undergraduate teaching across multiple disciplines. Through quite unusual interests of the then Provost, Phillip Griffiths-- a pure mathematician, who laudably developed the view that every leading university needed a core statistics department-- Duke launched a serious initiative to create a statistics centre, of some kind.  Important players were Bob Winkler (who had joined the business school at Duke in 1984), and external advisors including Jim Berger (Purdue) and Carl Morris (then at UT Austin) among an array of other senior leaders.  Economist and game theorist Roy Weintraub stood-in as interim director for ISDS, the Institute of Statistics \& Decision Sciences that was formally established in 1987, with mandates to \lq\lq deal with'' university-wide teaching of statistics and define a centre of excellence in statistics research.  A main theme-- this was mid-1980s, and nowadays it is perhaps not such a big deal as it was then-- was the outlook that ISDS would be an open-doors, interdisciplinary-oriented institute to \lq\lq get statistics out'' as well as  aiming to become a disciplinary centre of excellence.

\textbf{Hedibert and Filippo}: How was starting a group in a place that at the time was not among the leading statistics departments? How was the impact of passing from England to the US?

\textbf{Mike}: Well, Duke was certainly not then  ``among the leading'' institutions  in terms of presence of a statistics department at all! 
 I recall multiple conversations with others about my decision to move to Duke rather than other established departments; \lq\lq But there is nothing there!'' ... in statistics.  Of course there was a lot there in the mid-1980s, including research and educational excellence already engaged with statistics across many areas. Then, the clincher was my perception of the opportunities to help to define a new statistics centre, and to capitalise on the defined interests in doing things a little differently. This was consonant with my developing interests in moving Bayesian thinking forward by doing and engaging in applications, as well as by advancing core disciplinary R\&D.   I must add that
Jim Berger had major influence on my decision to \lq\lq visit'' Duke for a year; Jim and his wife Anne decided to join us at Duke for 1988/89.  My wife Lauren and I decided to extend the visit (!) while Jim and Anne came back permanently a few years later.

\begin{figure}[t!]
\centering 
\begin{minipage}{3.2in} 
\centering
\includegraphics[clip,width=2.5in]{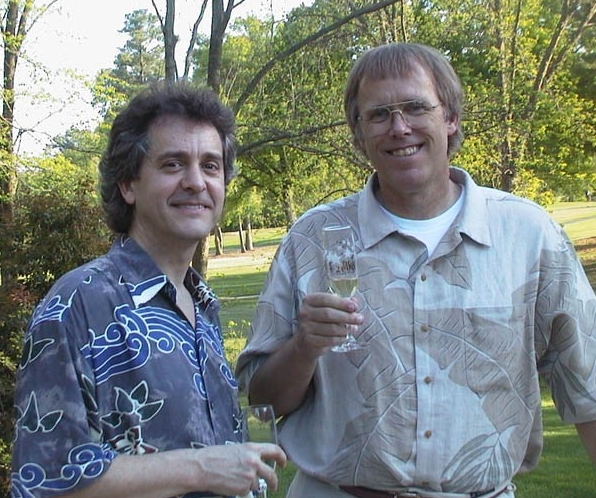}  
\caption*{With Jim Berger at Mike \& Lauren's home in Durham NC, 2000.
}
\end{minipage}
\end{figure}

In 1987, in the run-up to my moving to Duke, John Geweke, then professor in economics, took over as founding director of ISDS. John was and is a mover-and-shaker in Bayesian econometrics, who in those days was involved in major developments of importance sampling and allied ideas for Bayesian econometrics.  In tandem with helping to start ISDS, John was a key player in the initiative to establish the US National Institute of Statistical Science (NISS). While in transition from UK to US I became involved-- through John-- in this initiative, as well as in developing interdisciplinary grant activities at Duke.   The confluence of interests at the professional/international level in terms of the core rationale for the interdisciplinary statistics NISS initiative and the more focused university statistics institute at Duke  cannot be regarded as coincidental. This was a time of major evolution in thinking about the roles and relevance of statistics in society broadly ... and important in understanding my motivations and interests in taking a little \lq\lq time-out'' from UK and move to Durham NC.

In terms of my own research at the time, I had started a few new themes just before leaving the University of Warwick for Duke in 1988, and was in the final stages of drafting my first book-- Bayesian forecasting and dynamic models-- with Jeff Harrison, published in 1989. Among my new research areas were development of aspects of so-called agent (\lq\lq expert'') opinion analysis, and adaptive importance sampling for sequential particle filtering in dynamic models, each of which developed in the first couple of years at Duke with a series of publications in the early 1990s, and that have grown and continued to be main research areas.

\textbf{Hedibert and Filippo}: You became the director of ISDS (Institute of Statistics and Decision Sciences) in 1990, which would later become the Department of Statistical Science at Duke. How do you recall those first years? How was the statistics world at the time? 

\textbf{Mike}: It was an interesting and challenging time.  When I became Director after Geweke left, we had a small but, of course, immensely talented group: Michael Lavine who had been appointed from his PhD right at the start of ISDS,  Robert Wolpert and Don Burdick who had moved over from other Duke departments also at the start, Val Johnson who  had joined from his PhD in 1989/90,  and econometrician Jean-François Richard who then left Duke in 1991. Building up and out in statistics teaching across campus was a core goal, and we would not have done it without superb teaching from all involved, including an array of visitors and local, part-time instructors. Creating alliances with other departments and schools was fundamental; beyond working with departments on teaching developments, our interests and growing engagement in interdisciplinary research drove much of that. Important alliances also arose through establishing a formal consulting centre in ISDS, initially {\em pro bono} but then explicitly funded through the dean's offices across several schools. Active and vocal support from leaders and faculty in  schools of environmental sciences, engineering, business, areas of biomedicine, and aspects of the social sciences, were key. 

This was in an environment in which it has to be recognized that,  as for much of statistics in those days, faculty in several  academic departments-- and various significant political players in academia-- were fundamentally resistant and opposed to the idea of investing in a core statistics enterprise at all.   So, they were interesting and challenging times, and a learning experience for me on the academic-political side, in particular.  The drive to define a core research environment was, of course, fundamental and uniting, and defined perspective in investments. Initial progress on teaching and research fronts-- both core statistics and interdisciplinary collaborations-- quickly led to increased resources and substantial hiring of core faculty:  over $1991{-}1993$ this included Don Berry, Peter M\"{u}ller, Giovanni Parmigiani, Merlise Clyde, Dalene Stangl and Brani Vidakovic, with immediate impact across teaching and research, and swift growth and recognition of ISDS as a Bayesian and interdisciplinary research centre.

It was the (happenstance?) confluence of the interdisciplinary outlook in R\&D coupled with the core intellectual cohesion in statistical philosophy, and then at a time of immense advances in computation for Bayesian analysis, that helped it all come together ... well, to make it sort of work.  I will note that-- at the same time-- we were also helping to define and establish first programs at NISS, with the linked goals of \lq\lq getting statistics out'' in terms of broader engagement in interdisciplinary R\&D of all kinds. The 1991 appointment of Jerry Sacks as founding NISS director--  with a tenured faculty position in ISDS at Duke-- was a major development in that national initiative that helped to advance the infrastructure of statistical research, and the broader perception of the position and roles of statistics in society.

Those early 1990 few years were also personally immensely rewarding in terms of new connections and research collaborations, along with the remarkable inroads in core statistics research that our increasing faculty presence defined.   Then, the start-up process had a number of tolls, of course. One of note is that of starting a PhD program from scratch. It was 1993 before we graduated our first PhDs, and building to that at the start of ISDS meant that we lacked this core and vital component of a leading department. The following years expansion and eventual recognition have led me to state that it takes 10 years to establish a PhD program, 15 years to have it recognised as among the very best internationally. 

\newpage

\begin{figure}[htbp!]
\centering 
\begin{minipage}{3.2in} 
\centering
\includegraphics[clip,width=3.0in]{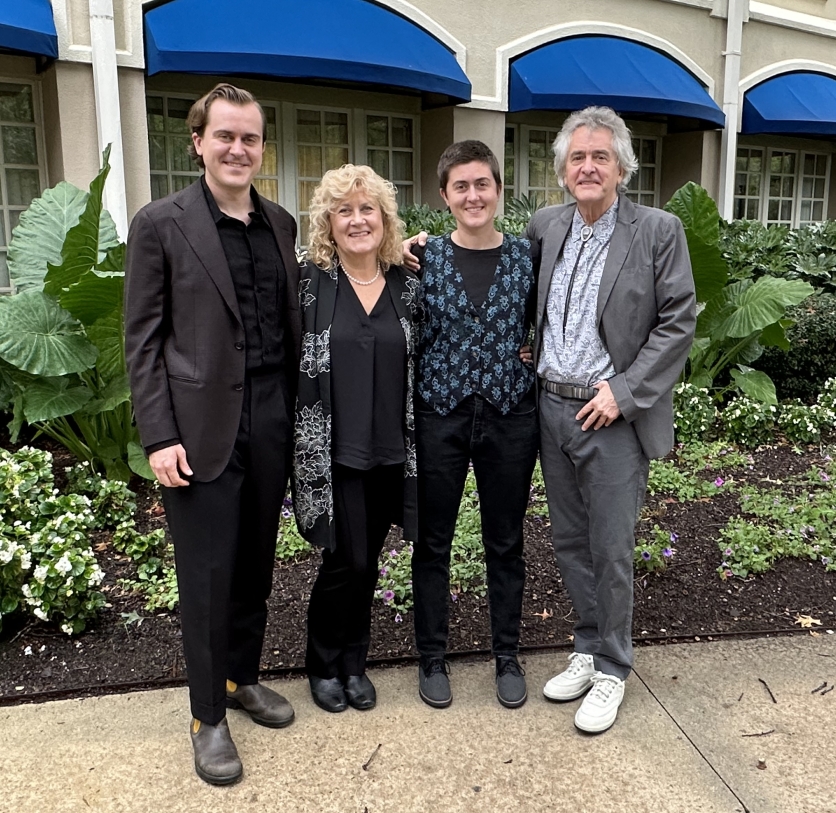}  
\caption*{Dylan, Lauren,  Abagael and Mike at a Duke departmental dinner celebrating Mike's move to Emeritus status, September  2024.
}
\end{minipage}
\end{figure}

The same few early years in USA were also immensely important and rewarding in personal and family life.  Lauren and I welcomed our daughter, Abagael, in 1989 and son, Dylan, in 1993. 

\textbf{Hedibert and Filippo}: Duke's Statistical Science department has, for some years now, been recognized as one of the leading groups especially in Bayesian statistics. A lot of this fame comes from a group of researchers (you, Jim Berger, Merlise Clyde, Robert Wolpert, Alan Gelfand and many others) who spent much of their academic career at Duke. How has this remarkable group been created? Why do you think you all decided to stay?
You surely had a tremendous impact on the development of this department. How do you think, vice-versa, that your career has been shaped by belonging to this group of people?

\textbf{Mike}: The main factors are and have always been those of intellectual community and engagement with common goals for the department, institution and the statistical profession more broadly.  The expansion of foundational and applied research involving Bayesian statistics-- and the integration of Bayesian ideas and methods in teaching at all levels-- have been distinguishing features of the Department of Statistical Science as they were in the establishment 
of ISDS.\footnote{After developing as a free-standing Duke institute and launching the PhD program,  ISDS became a department of Duke's School of Arts \& Sciences  in 1990. We maintained the name and \lq\lq brand image'' of ISDS until 2007 when, with our establishment of the undergraduate major and minor in statistical science, we changed to the Department of Statistical Science.  For brevity, the new name dropped the \lq\lq decision'' that was key in ISDS  in reflecting the initial goals and  part of the rationale for its creation in the late 1980s. I regard the scope of Statistical Science  as encompassing decision theory and decision analysis-- the \lq\lq Yang'' of our discipline that complements 
the  \lq\lq Yin'' of probability, modelling, data, inference and prediction.   That said, when I was awarded distinguished professorial status at Duke University in 1999, I deliberately chose the title of my chair to include \lq\lq Statistics \& Decision Sciences'', recognising and maintaining the historical tie to ISDS. On a related point, I always thought that we should include \lq\lq Art'' in the departmental name, since much of what we do is creative art in terms of exploiting imagination and skills to define new ways of thinking and looking at the world through statistics. I believe that, in the public eye, statisticians are often under-regarded as core members of the so-called \lq\lq creative class''.}

Since the early years this set Duke a little apart from other leading universities; this  was and has since been a main attraction for students and faculty alike.   The coupling of core Bayesian interests with main advances in computation in the 1990s bred a fertile environment for methodological research, and this critically played into the rich expansion of interdisciplinary collaborations.  The latter became a hallmark of the department with vibrant collaborations across multiple schools at Duke and internationally, and with NISS and then later through the 2002 establishment and activities of SAMSI (Statistical and Applied Mathematical Sciences Institute, led by Jim Berger) over two decades.   This intersection of  personal academic and research interests with engagement in broader professional development has been important in terms of faculty coming to, and staying at, Duke.   It is also important to recognise the ways in which faculty moving to other institutions feeds positively into these broader interests.  Over the years, a number of Duke faculty have moved onto other institutions-- including several past or current chairs of leading departments, and key players in developments of new programs at others. 
Then, very naturally, faculty interactions on day-to-day and longer-term bases have enormously influenced my own thinking and statistics in research, as they have in impacting my engagements in professional developments more broadly.   In terms of \lq\lq shaping'' my career, surely continued interactions with the faculty we attracted to Duke over the years, and their engagement in leading and promoting developments in 
Bayesian statistics broadly and longer-term,  are main factors in why I have been here for 37 years. 

Personal and family circumstances factor in centrally, of course.   Duke in Durham and the Research Triangle Park is an attractive 
environment as a professional socio-economic \lq\lq hub'' and for personal and family lifestyle considerations.   The population has expanded dramatically since 1988 (the RTP region has more than doubled in population in the last 30 years, and continues to grow 
apace with current population well over 2 million), but still 
does not suffer so much from the day-to-day constraints of large cities.  Statistical science faculty have always tended to live close to the university and this has been really important in terms of life balance and for community building. Part of the continued success of the department is that people are and feel-connected outside of academics, in personal and family life. We have generally thrived on development of friendships \lq\lq outside business hours''-- faculty, postdocs and PhD students alike-- friendships that help to engender intellectual community and feed-into productive collaborations as well as contributing to broader satisfaction with the professional environment.

\textbf{Hedibert and Filippo}: Opening a very serious part: Halloween parties were always a big moment for the department. How did this start? Any episodes you want to recall?

\textbf{Mike}: My wife Lauren and I had always enjoyed Halloween and had a series of parties over the years in England, prior to moving to Durham
(somewhat coincidentally, my birthday happens to be October 30th).
We had a break for a couple of the initial years at Duke-- just the transition and family circumstances, the latter including the arrival of our first child, daughter Abagael in 1989.    Later,  I promoted the idea within our faculty. Not just that I really enjoy Halloween, but that such \lq\lq departmental'' gatherings can be useful ice-breakers, especially for welcoming newcomers to the department. 
 We kicked-off the Duke event in 1992,   hosted by then newly appointed assistant professor Giovanni Parmigiani. For a number of years we hosted the event at a faculty home, until the department outgrew that.  The series has been running ever since (with some virtual or hybrid versions in pandemic years 2020/21).  
On the (really) serious side, this connects with my above comments about the importance of the non-academic socio-professional interactions and community-building.  The Halloween parties provide timely get-togethers for faculty, postdocs, PhD students and visitors mid-way through the first semesters of the academic years.  Other departmental events through the year are as important, but the timing of Halloween is particularly important for new/incoming junior researchers and especially the 1st year PhD students.  After what is always a busy and sometimes frenetic few weeks at the start of the academic year, students get to see faculty in a different context, and one that can and often does contribute to easing communication and engendering colleagueship thereafter.  Many faculty members are professors because they enjoyed being PhD students; new PhD students are often unaware that professors are students too.   Then, these are also family events, and some of the more creative costumes are often those of the children of members of the department. 

On the lighter side, there are of course many \lq\lq episodes'' that I could mention, and many photos. I recall them all, but here will mention 
the 1996 gathering, my 40th birthday: the PhD students colluded and many of them turned out costumed as (masked versions of) Mike West.  An example of the creativity of students. It has been said that Mike promoted the Halloween parties as a way to \lq\lq filter'' incoming PhD students according to creativity in costumes and their stagecraft ({\em an imaginative costume is required!}); there might be sometime to that. Another vignette comes from 2016, when the department had several visitors from Europe, including a number of Italian academics. In collusion with some of our Italian and other EU PhD students, they presented as anti-Brexiteers.  It happened that my wife Lauren showed up as Britannia (this was tongue-in-cheek; Lauren was not, shall I say, a supporter of Brexit!)-- a perfect counterpoint.   Then, I showed up as the British Electorate.  The Italians, in particular, somehow missed my point-- I was there representing the almost 50:50 split of the electorate with my costume made in equal parts from  EU and UK flags (our Halloween outfits are always home-made).  Some other memorable (for me) episodes relate to karaoke and related performances of various members of the department. Apart from the \lq\lq fun'', this can be informative in illuminating innate personal creativity and, again, stagecraft of colleagues and students-- characteristics fundamental to success in academia. 

\begin{figure}[htbp!]
\centering 
\begin{minipage}{3.2in} 
\centering
\includegraphics[clip,width=3.2in]{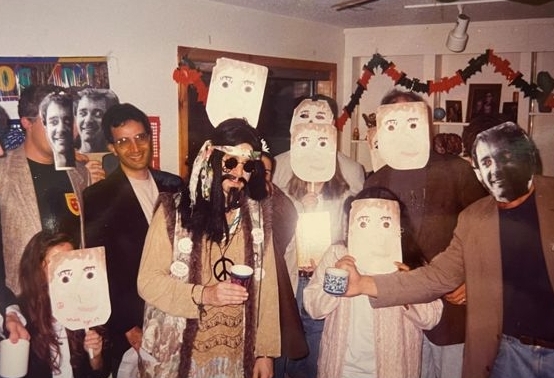}  
\caption*{Several \lq\lq Mikes'' at the ISDS Halloween party 1996, coincident with Mike's 40th birthday. 
}
\end{minipage}
\hskip.25in
\begin{minipage}{3.2in} 
\centering
\includegraphics[clip,width=1.4in]{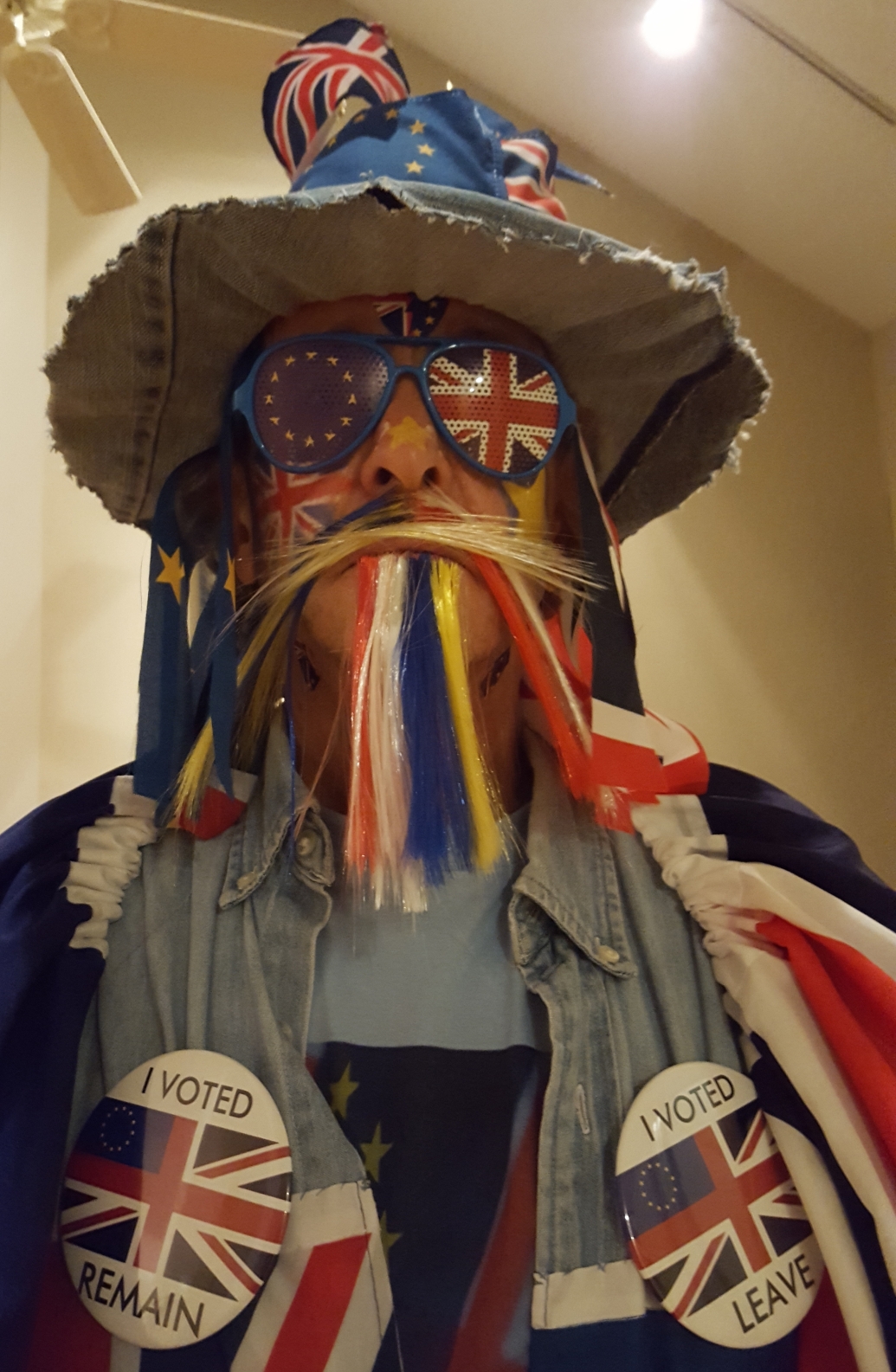}  
\caption*{Mike as the British electorate, Halloween 2016.
}
\end{minipage}
\end{figure}

\textbf{Hedibert and Filippo}: We see 45$\pm$ PhD students, including current ones, and more than 20 postdocs, to date. How did you decide on your style of supervision?

\textbf{Mike}: I can't say I have ever thought so much strategically in teaching and mentoring, including PhD student advising and postdoc mentoring.  I do like to use the term {\em advising} rather than {\em supervising}, and that reflects my experiences with PhD students, in particular, over the years. I have no examples of PhD students collaborating with me where I defined a specific, \lq\lq do this then that .. and a PhD will emerge'' project, and supervised it.  Rather, I have enjoyed working with and collaborating with many PhD students, typically starting out with really informal conversations about areas of interest and importance-- in my view and in their view-- that intersect with our current interests. And then advising by responding, coaxing, nudging the students' investigations, helping to keep things on track in terms of progress on core PhD research as well as-- in many cases-- linked projects that they and I are involved in with collaborators.  Always ready to step-in with a somewhat heavier nudge from time to time as current directions seem to be-- perhaps-- less propitious than others I can imagine; but always open and expectant that the student's self-directed tracks into research will lead to new ideas and relevant advances that I would not naturally conceive.   Then, a core and critical aspect of mentoring-- PhD or otherwise-- is engagement with the broader professional development of the mentoree.  Focused advice and assistance in building out into the professional community-- promoting and helping with internships, conference participation, and other broader professional interactions--  has always been important in my PhD advising as it should, in my view be, for all PhD advisors.

\begin{figure}[b!]
\centering
\begin{minipage}{3.2in} 
\centering
\includegraphics[clip,width=3.4in]{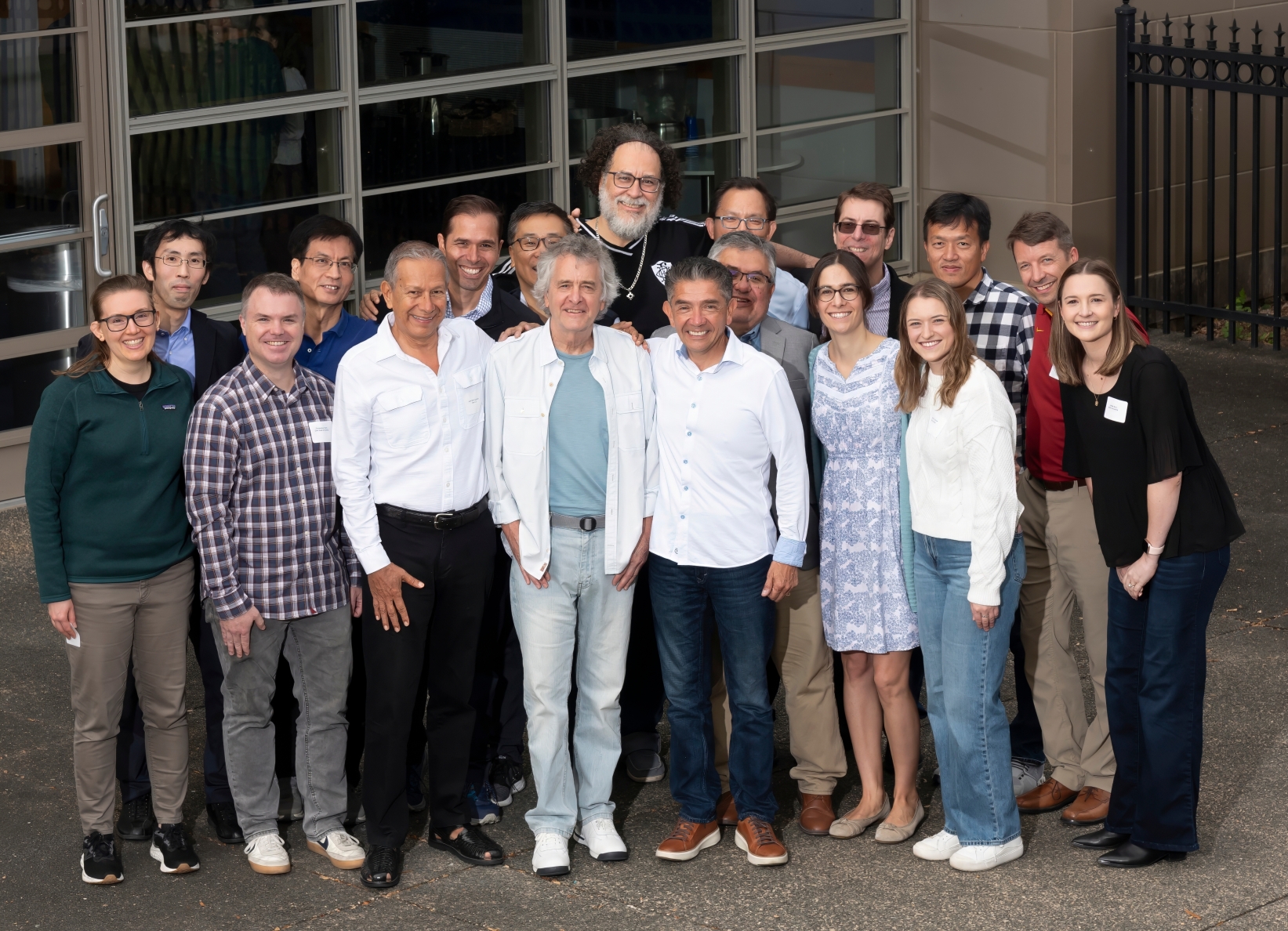}  
\caption*{Mike with a ${\sim}25\%$ sample of past PhD advisees and postdocs on \lq\lq Mike Day'' at the Statistical Science Research Alumni Symposium at Duke University, November 2024.   
}
\end{minipage}
\end{figure}

 I have had a few PhD student  collaborators who approached me with some specific ideas in mind, and they then generally moved fast to PhD research. Usually, however, it has been a matter of learning collaboratively and exploring new directions as thinking evolves.   I will say that, at the outset in conversations with new and aspiring PhD students-- and new and aspiring researchers more generally-- I have been fond of asking a hard question up-front:  What is research?  I recall asking Adrian Smith a related but   different question before I started PhD; my question was about \lq\lq the mechanics'' of doing a PhD. What does a PhD student in statistics do every day? It is a hard question, and was harder in the late 1970s/early 1980s when the answers were limited relative to current times. Nowadays, the engagement in ranges of projects, from very applied to more esoteric, is common across PhD programs in statistics, and there is a lot to do that is overt and obvious: on top of the core \lq\lq read and learn'' from the literature and through interactions with other students and faculty (that was basically \lq\lq it'' in the late 1970s),  there is coding, data analysis, reanalyses, learning by doing with exploration of models and methods across ranges of contexts, and the ability to engage in this broad sweep of activities that underpin modern statistical science research.    Then, returning to my \lq\lq hard'' question: research is about {\em asking questions}. Asking new questions, rephrasing existing questions, convincing oneself and others that the questions are interesting and important, presenting instantiations of the questions to push forward that argument, and then coming up with potential solution paths and maybe/sometimes  solutions. To circle back on \lq\lq my style of advising'': whether with PhD students, or students at other levels, or collaborators in research, progress relies so fundamentally on innate curiosity and the knack of asking the \lq\lq right'' questions.  \lq\lq Thinking out loud'' and asking 
\lq\lq silly questions'' in group discussions, seminars and other forums is something I grew to be fond of years ago, and have found more and more useful over the years in advising new researchers. Lead by example:  ask silly questions in public, and encourage others to do so.

Successful and productive professional collaborations rely on growing inter-personal relationships. Defining and developing new research collaborations of any kind does so,  and my own experiences with PhD students and postdocs have reflected this. Students and postdocs work with advisors for a number of years, and for them it is a huge and dominating focus, typically at a time of life of major change and personal development. I have been fortunate to have been advisor for many wonderful statisticians, most of whom became personal professional friends very quickly; I doubt that the professional collaborations would  have worked so well otherwise.  Many and most also became personal and life-long family friends. While not part of any strategy or planned \lq\lq style'' of PhD advising,  this reflects the importance of developing one-on-one personal relationships that are simply fundamental to collaborative research and the health of the 
research community.

\section{Connection with other groups}

\textbf{Hedibert and Filippo}: You contributed to the development of Bayesian statistics all over the world. How did this happen? For example in Brazil...

\textbf{Mike}: When I started at Warwick University, Helio Migon (from Brazil) was one of only 2 PhD students there, working with Jeff Harrison; 
the Brazilian connection had been initiated a few years earlier with Reinaldo Souza, who finished his PhD with Jeff in 1979.   Just after my move to Warwick, 
Dani Gamerman (Brazil) and Pepe Quintana (Mexico) arrived for PhD and became my first two PhD advisees.   There were lots of interactions among that small group that led to collaborative research over a number of following years, and to lifelong friendships.  Helio and Dani moved back to Brazil to help to build the research environment in Rio,  and our continued interactions after I moved to Duke helped to fertilise activities.  Our new PhD program was fortunate to attract really top students from Brazil and Mexico-- as from USA, Italy and China-- in the early years.  On the  Brazilian connections in particular, 
we have too many Duke PhDs to name here, but I have to mention my own past advisees Hedibert Lopes, Marco Ferreira, Carlos Carvalho and Fernando Bonassi. A number of our Duke PhD graduates have moved back to Brazil over the years to contribute to the development of graduate programs at a number of universities (UFRJ, INSPER in S\~ao Paulo, UFMG among them) and have developed as international professional leaders in our societies as well as research leaders. 

\textbf{Hedibert and Filippo}: Italy, ...

\textbf{Mike}: My personal connections with the Italian Bayesian community grew through engagement in international conferences in the early 1980s and then with visitors the Warwick Bayesian Study Year. There was of course a strong history of  Bayesian interests in universities in Rome, Milan and Pavia, in particular;  some tied back to de Finetti, of course, and in the early 1980s with Eugenio Reggazini and others at Bocconi University among other Italian groups. In those days, some active areas linked across 
 Bayesian foundations, reliability/survival analysis and intersections of stochastic process theory with what was then neophytic Bayesian nonparametrics.  
On establishing the PhD program at Duke, our then new assistant professor Giovanni Parmigiani was a live connection to Italian academics, and played a role in attracting superb PhD students to Duke in those early years. Among the very first few in the early 1990s,  Claudia Tebaldi and Giovanni Petris both arrived from Bocconi (and I happily agreed to advise them for PhD), while Fabrizio Ruggeri (who was advised by Michael Lavine) arrived from Milan via his MS at Carnegie-Mellon.  This developed as a main \lq\lq pipeline'' of Italian students--  too many to name here-- that continues (and I hope and expect will continue to flourish). One main development over the last two decades been the series of workshops on Bayesian Inference in Stochastic Processes (BISP), established and largely run by Fabrizio.  These have been important meetings in the infrastructure of research in Bayesian analysis, and-- while open, international conferences-- have maintained strong representation of Italian and Duke researchers and students over the years.   

\textbf{Hedibert and Filippo}: Mexico, ...

\textbf{Mike}: Connections with faculty in Mexico were most relevant in the early years of ISDS, particularly at the Autonomous Technological Institute of Mexico (ITAM) and the National Autonomous University of Mexico (UNAM).  This led to a number 
of really superb PhD students coming to Duke, some of whom now stand as major professional and intellectual leaders as well as personal and institutional friends. Too many to mention here, but I will note the first arrivals to our new PhD program at Duke in 1990s-- Omar Aguilar (Schwab Investment Management), Gabriel Huerta (Sandia National Laboratories) and then Viridiana Lourdes-- who I was fortunate to advise and who are all lifelong friends. 

\newpage

\textbf{Hedibert and Filippo}: and Japan.

\textbf{Mike}: I have enjoyed and been engaged in interactions with Bayesians in Japan since the early 1990s.  
This was initially and mainly associated with the national Institute of Statistical Mathematics (ISM) through Hirotugu Akaike and Genshiro Kitagawa, and later with research leaders from that environment, with natural links to interests in time series and state-space modelling, in particular. Over the years this developed more broadly with institutional advisory roles and connections, Japanese researchers spending time at Duke, and then with a flow of Japanese students joining our PhD program.  Two of my past PhD advisees and friends-- Jouchi Nakajima (Hitotsubashi University) and 
Kaoru Irie (University of Tokyo)-- are  in senior positions in academia in Japan, as are a number of other Duke PhD alumni. This helps to continue to 
grow statistical graduate education-- with a Bayesian flavour, among other things-- in Japan, and 
I look forward to continued and increased flow of intellect both ways.

\begin{figure}[htbp!]
\centering 
\begin{minipage}{3.2in} 
\centering
\includegraphics[clip,width=2.2in]{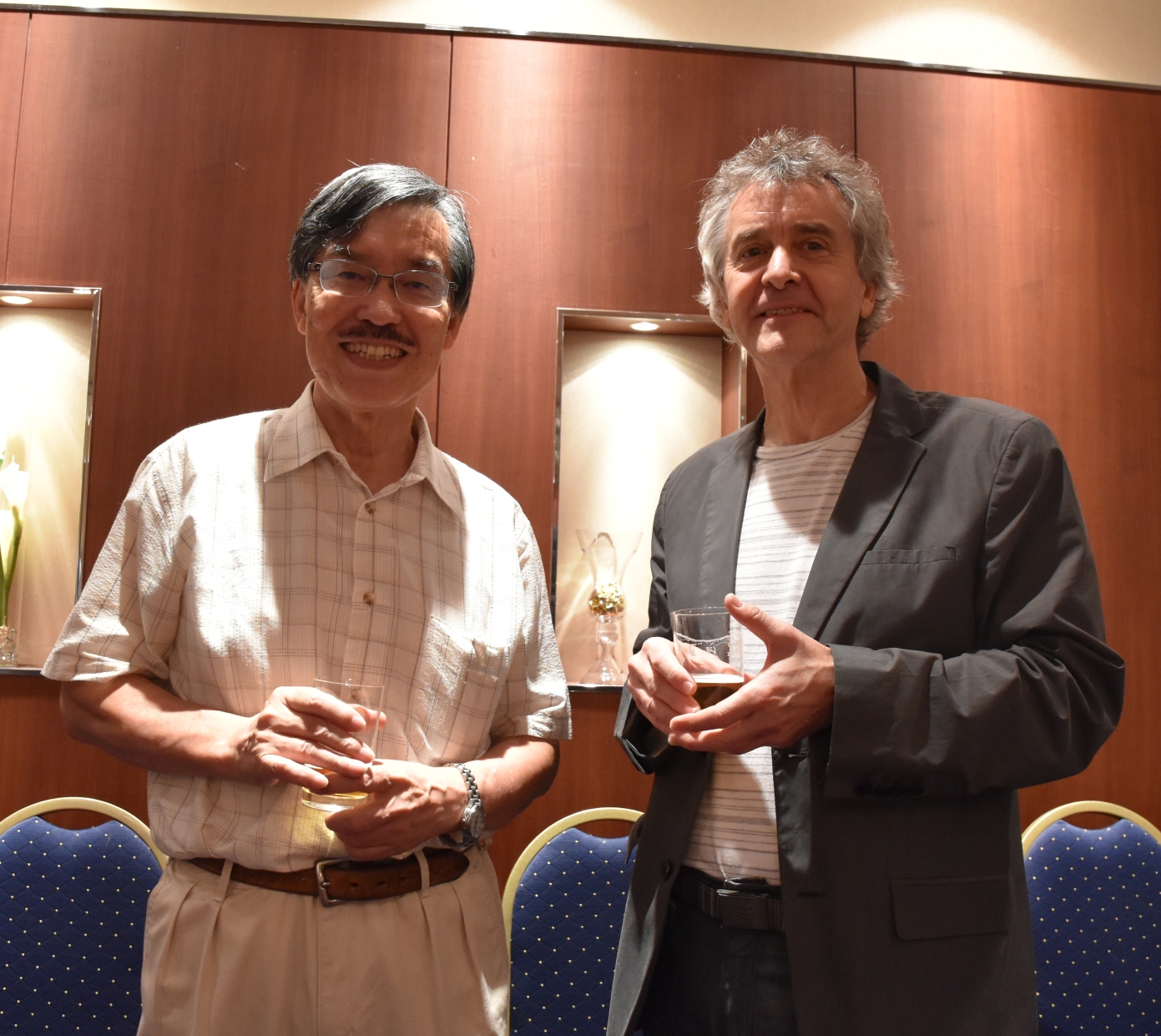}  
\caption*{Genshiro Kitagawa (past Director-General of the Institute of Statistical Mathematics, Japan) and Mike in Tokyo, on the occasion of 
Mike receiving the Akaike Memorial Lecture Award in 2018. 
}
\end{minipage}
\end{figure}

A final note on this general theme is that, over some years following the establishment and recognition of ISDS in the 1990s, a number of US universities interested in either reinvigorating or starting a statistics department or centre-- of some kind-- did find interest in consulting with us about our experiences-- and the \lq\lq why, what and how'' of  our institute and then department at Duke. The \lq\lq model'' was rather new and different, and the interdisciplinary outlook (and success) garnered interest at institutional levels as well as in the statistical profession. I could mention a number of institutions that \lq\lq came calling'' over the years and with whom we engaged and in most cases, I believe,  contributed productively.  I will  
specifically note the Departments of Statistics, at  the University of California at Santa Cruz, and  of Statistics \& Data Sciences, at the University of Texas at Austin.  These and other universities found some useful data and insights in discussions of our experience at Duke, and a number of our PhD and postdoc alums-- and friends-- that moved there have been central to the establishment, growth and current vitality of their programs.

Our professional community is very connected geographically, and the development of Bayesian statistics in many regions of the world has thrived through conference interactions and the movement of PhD students and postdocs, in particular.  I have been fortunate to teach and mentor students from across the USA and from many countries, and part of the richness of PhD activity has come through the diversity of students' backgrounds.   Graduating students and postdocs 
moving elsewhere-- not only to academia in USA and worldwide-- have been and are fundamental to our profession (as well as specifically to the reverse-fertilisation with new PhD students coming to Duke!) This has been most important for the development of the Bayesian community worldwide over the last $3+$ decades.

\textbf{Hedibert and Filippo}: You were the elected  president of the International Society for Bayesian Analysis (ISBA) in 2009-2010 (the foundation of ISBA itself is also very connected to Duke). What do you think about the development of the society over the years? And similarly for the Section on Bayesian Statistical Science of the American Statistical Association (SBSS). 

\textbf{Mike}: The community motivations for creating ISBA, a free-standing society, were consonant with those that led to SBSS,  but more expansive 
 in terms of emphasising and promoting international and interdisciplinary concerns.  ISBA is nowadays the world's central professional body representing Bayesian analysis broadly, and has evolved to a position of prominence   through international conference organisation and the highly-ranked {\em Bayesian Analysis} journal, easily my favourite statistics journal. The array of conferences organised or sponsored by ISBA and its Sections are major contributors to our profession, and the biennial World Meetings are highly regarded.   In 2000, I was program chair of the ISBA World Meeting in Crete and also a member of the organising committee for the series of Valencia International Meetings that had been running since 1979. The 2000 meeting was a break-out for ISBA in terms of attendance,  and at that meeting we made the decision  to \lq\lq merge'' the Valencia meetings with ISBA World Meetings, the latter subsuming the former a few years later. Understanding the importance of the Valencia series to the growth and presence of Bayesian statistics  in the 1980s/90s, the merging with ISBA represented the growing recognition of the position and relevance of a formal, free-standing society albeit only 8 years since establishment.  The ISBA World Meetings are very successful (over 800 participants in Venice in 2024) and of course look set for continued growth in coming years.  One main challenge for all free-standing societies is to maintain growth of active membership; this comes through clear focus on what the society does for the members.   One of my first  actions as ISBA President in 2009 was to establish life membership (and I was the first to sign-up!) with a view to encouraging longer-term engagement. A key point was that the life membership fees contribute to a fund to support junior researchers. The ISBA conferences and workshops stimulate membership, of course, and the growth of ISBA Section conferences is very healthy. Some of the Sections-- including the vibrant j-ISBA (Junior-ISBA, new researchers) Section-- have their own,  successful conference series, as do some of the ISBA Chapters representing regions around the world.
 
I will add that ISBA has and continues to very positively reflect the diversity-- geographic,  generational and scientific diversity included-- of individuals engaged in Bayesian R\&D and education.  One of the founding goals of ISBA was to promote interactions of statisticians with disciplinary scientists through collaborations with other societies.  Much was done in the early years to connect with, for example, physicists through MaxEnt and economists through the growing activity in Bayesian econometrics. Over the years, formal activities related to this aspect of ISBA's founding vision have, perhaps, been less evident.  This might be regarded as an area for focus for future efforts of ISBA, especially in view of the growth and development of Bayesian methods in many disciplines in the last two decades. That said, the imperative to \lq\lq get Bayes out'' into applications is much less keen these days; while ISBA has played key roles in that, it also has happened organically as the discipline has flourished.

\section{Research}

\textbf{Hedibert and Filippo}: How would you describe your research?  How do you choose the questions of interest?

\textbf{Mike}: I am fond of letting students know my view that, for many, PhD study is  the best time of their professional lives-- you get to do anything you want in research. That's been my view and experience as a life-long student.  An element of strategic thinking about how much time and intellectual energy (and sometimes money) to devote to a particular line of research among many has been relevant at times, but much of my research has evolved mainly by following my interests and those of collaborators, especially students.  I have never been much of a \lq\lq follower of fashion'' in terms of jumping (at least not so quickly) on new and fashionable topics, and core methodological themes in my research have persisted over the decades reflecting my interests. As I noted earlier, the interplay between various applications motivating new methodology that then applies elsewhere has been, for me, a very fertile interplay; that is and should be a main driver of much relevant theory and methods in statistics more broadly. 

Some vignettes are drawn from elements of my research during the 1990s.  I was very involved in developing aspects of the methodology of dynamic models, particularly with respect to exploring implications for underlying structure in univariate and multivariate time series.  Part of that was motivated by applications in climatology (partly with geologist and climatologist Tom Johnson)-- where inference on underlying trends and quasi-periodicities, in particular,  are key. As this developed I also  became engaged in collaborations with experimental neuroscientists (especially clinical neurophysiologist Andrew Krystal) involving EEG data, and the same model forms and methodology became central to that area. Stepping-back from specific applications led to a deeper dive into latent factor models and their time-varying extensions, with fruitful applications in financial time series. In terms of continuing core statistical research, this also then evolved to explore early versions of sparse latent factor models and dynamic extensions for time series. These are nowadays broad areas of methodology and active research-- in both core statistics and   interdisciplinary applications.

\textbf{Hedibert and Filippo}: So a lot of your works originated from conversations with experts outside statistics.

\textbf{Mike}: Happenstance meetings that lead to collaborations can be important in guiding core statistical research. This has been notable for me in a number of areas, including the two just mentioned above.   I will add just two more here.  First, right at the start of the 1990s, a chance and really routine consulting conversation with neurobiologist   Dennis Turner led to a substantial collaboration involving development and applications of structured mixture models. I had, of course, been interested in mixture modelling in various aspects since my PhD, including exploring  foundational aspects of mixtures to try to underpin methodology for \lq\lq smoothing Monte Carlo samples'' in Bayesian analyses, both static and in sequential dynamic contexts.  Developing customised models for the specific biological contexts (concerning fundamental biochemical mechanisms of synaptic  communication) led to broader research on mixture model methodology that then interfaced with my other interests in mixtures.

The other vignette and example I will relate here--  far-reaching in terms of its positive impacts on my career in many ways--   began with a fairly random 1999 phone call with leading molecular geneticist Joe Nevins.  He had asked to chat about some statistical problems, and we arranged a phone call while I was away at a conference. Of course, Joe had an agenda and a grant deadline coming up.  The \lq\lq statistical problems'' had to do with then-new gene expression microarray technology.   One main reason I was interested is that the multivariate dynamic factor models for time series that I had been developing in other areas seemed just right for a potential new and important application: the conversation was about  much higher-dimensional gene expression time series in the human cell cycle setting, with data generated under a carefully designed multi-factor (several environmental conditions, genetic interventions, treatments). Well, that random conversation led to very fruitful collaborations for more than a decade. We did not generate that 1999-conceived data set; the initial plan was simply experimentally and economically infeasible in those days, and would still be very challenging now. We did, however, develop a much smaller study and data set that led to one of the very first microarray-based publications (in 2001) concerned with the human cell cycle.  That was just 
one initial result of the happenstance conversation that led to new \lq\lq questions of interests'' and statistical research in multiple areas:   fundamental biological and technological questions of  data quality and calibration (and the founding of a biotechnology company-- one of the first in the USA to receive Federal approval for providing data from gene expression microarrays for diagnostic testing); developments of large-scale, sparse factor models-- including the introduction of the term \lq\lq large $p$, small $n$'' (which, as Geoffrey Grimmett pointed out to me years ago, should of course have been \lq\lq large $P$, small $n$'')-- first in my talk at a November 2000  workshop on statistics in functional genomics at the Institute of Pure and Applied Mathematics, a.k.a. IPAM at UCLA, and then eventually published in 2003; parallel developments of Bayesian methodology for large-scale graphical models; statistical modelling for formal cross-context and cross-study data integration and inference (presaging by a decade or more what is nowadays referred to as  \lq\lq transfer learning''); and then developments into mechanistic aspects of systems biology coming back, in part, to my interests in sequential, time series problems.  So this particular interdisciplinary collaboration was arguably unique in stimulating ideas and directions for new, basic statistical research of much broader interest, and allied professional activities.    

I have been fortunate to have had a series of initial interactions  with truly exceptional disciplinary scientists that have led to detailed and sustained collaborations that had such broader impact.  On paths ahead in thinking about how to choose and define new research directions in core statistical theory and methods, I believe we should be open to influences that are often unpredictable, and potentially profoundly influential. I have more, and more recent examples in terms of my own research, touched on below.

\textbf{Hedibert and Filippo}: Among your biggest contributions there are surely the ones related to dynamic linear models. How did you get interested in them in the first place?  

\textbf{Mike}: As I noted earlier, I developed some interest in time series as well as Bayesian analysis in my final undergraduate year.   When I started my PhD with Adrian Smith, one of the topics he was interested in linked to the recent developments in Bayesian forecasting, and specifically to the-- then recent--  methodological  advances in state-space modelling of Jeff Harrison and collaborators. Connecting to more traditional time series approaches that I had seen with Paul Newbold, this seemed at least as interesting as some of the other areas we discussed (and who would not want to engage in \lq\lq dynamic'' PhD research?). As a seminal innovator and pioneer of hierarchical modelling and Bayesian methods, this was a natural next-step for Adrian;  Jeff Harrison's formulation of dynamic models (dynamic linear models, to begin) with Colin Stevens and others defined state-space representations that are really just that-- extensions to sequential, time-varying settings of hierarchically structured, random effects models. Then, of course, with a lot of smoothing over time induced by careful model structuring and prior specification. The framework addressed local and global non-stationarities, included and encompassed practically all of the existing (linear) time series models that we knew at the time, and opened new paths to generalisations.  The themes embraced forecasting (a.k.a. prediction) as a core goal and emphasised the subjective Bayesian perspectives of admitting various forms of information into analyses. I was sold on this as a rather new and \lq\lq dynamic" research area from my first readings at the start of my PhD years.

The roles of Dennis Lindley in helping with perspectives in those initial years were important. It is to note that, in addition to being Adrian's formal PhD advisor, Dennis was also an influential   mentor/advisor  of Jeff Harrison at Cambridge in 1958/1959; Dennis later played a key role in Jeff's move from industry to found the Statistics Department at the then-new University of Warwick in 1972.  In that sense,  
 Jeff was one of a number of my adopted \lq\lq intellectual uncles'' in terms of academic genealogy. 
(As an aside, one of my active non-statistical interests for many years has been genealogy-- I am the keeper, primary researcher and maintainer of my family tree).   
As I started my PhD, learning  a little more about the preceding developments in the professional culture of statistics, and having some initial associations with key people and personalities involved as well as \lq\lq insights'' into the frontiers of R\&D, were ingredients in my thinking about where to spend time, and on what.   
 
\begin{figure}[htbp!]
\centering 
\begin{minipage}{3.4in} 
\centering
\includegraphics[clip,width=3.4in]{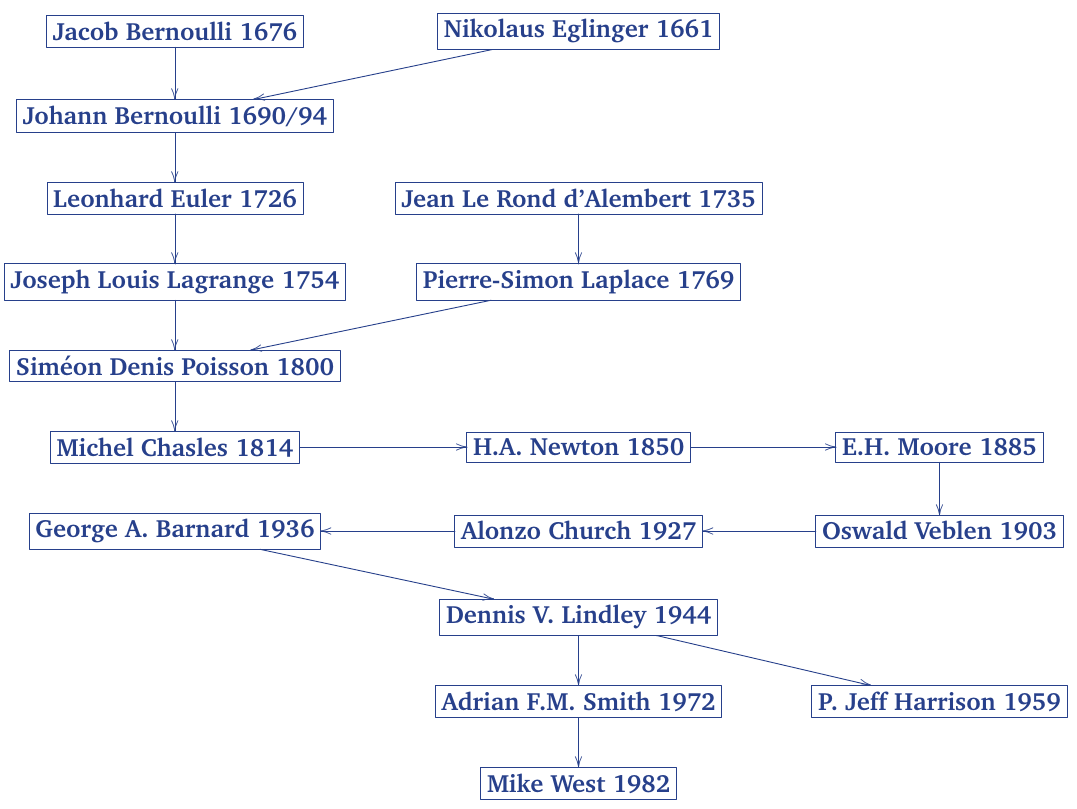}  
\caption*{Mike's Intellectual genealogical tree (some data from www.mathgenealogy.org) with dates of main degrees or notable events in advising relationships.
\label{fig:mwgenealogy}} 
\end{minipage}
\end{figure}

\newpage

 \textbf{Hedibert and Filippo}:  Your most cited work is the book on {\em Bayesian Forecasting and Dynamic Models}. What was the genesis of that book? 

\textbf{Mike}: In my first year on the faculty at Warwick (starting in October 1981 when I was still writing my PhD thesis) I taught two courses-- one on multivariate statistics and one on time series and forecasting.   New assistant professors know that the first year-- and first few years-- of teaching are when and where we learn the basics; my experience was no different.    Then, the opportunity to teach an advanced undergraduate course in time series and forecasting was just what I wanted and needed. Through that and the following year, and with routine interactions with Jeff Harrison on ranges of relevant topics, the \lq\lq need'' to teach coherently and comprehensively-- with teaching involving relevant and engaging applied examples-- was a key and critical learning period. It was also a time of recognition that figuring out how to teach can define important stimuli to research, as well as building early engagement with the realities of communication to heterogeneous audiences. One result was a main theme in research in time series forecasting and monitoring; another, in parallel, was the development of material that built on the existing foundation Jeff had defined, and putting it together in book form was then a natural next step.  Well, starting a book intended to be a comprehensive research monograph generated a major concern for theoretical depth and detail as well as methodological breadth.  Then, as it developed it naturally spawned new ideas and  a collection of new research explorations and multiple new papers.
So, it took a few more years! The main material was in place by late 1987, but then distractions created partly by my move to Duke in 1988 delayed the publication to 1989.
 I will say that a critical part of our development over those early-mid 1980 years was software for applications. A main stimulus on that count was early industry research collaborations (with companies including Imperial Chemicals Industries, a.k.a. ICI, on demand forecasting at multiple scales; forays with IBM and other companies on regional revenue forecasting;  and  with market research companies on understanding consumer responses to marketing campaign interventions). I wrote the first version of BATS-- Bayesian Analysis of Time Series-- in just three months in fall 1984; that  first version was programmed in Basic on one of our first departmental IBM PCs. In the following two years, I morphed this into APL. These developments were important in contributing real-world examples to the first edition of our book in 1989. Later,  our postdoc,  collaborator and then family friend Andy Pole defined the C implementations for our second book, {\em  Applied Bayesian Forecasting \& Time Series Analysis}, that appeared  in 1994. Largely through Andy's contributions, we released the software with the book; this was at a time when the questions of access to implementations, and the entire enterprise of reproducibility of statistical analyses, were still on the road to becoming mainstream.

\textbf{Hedibert and Filippo}: Your 1995 paper on Bayesian inference using mixtures is your second most cited work. When did you get interested in nonparametric methods?
 The algorithm you describe is now the cornerstone of many BNP analysis. How did you arrive to it?

\textbf{Mike}: I have been interested in various aspects of  mixture distributions and mixture modelling since my PhD days.  My research in the late 1980s/early 1990s touched on  a range of emerging topics, including the use of mixtures as importance sampling distributions in both static and sequential, dynamic settings. I exploited kernel and clustering ideas in some of this, but was then concerned about the lack of  more theoretically-based Bayesian ideas for the problems of  \lq\lq smoothing Monte Carlo samples''. This led me (naturally, in retrospect!) to the (then, not so extensive) theoretical literature on Dirichlet process mixtures. By the middle of 1990 I had defined a computational method that explored large numbers of so-called configurations (clusters) of samples and then used their theoretically defined weighted averages as   approximations to predictive densities in such models. I was pleased with this as it had a foundational basis for otherwise {\em ad hoc} clustering methods. I presented this a couple of times but never got around to writing it up as a paper; the reason for that is that I met Michael Escobar,  and he was ahead of me (and everyone else) in bringing what we later called Markov chain Monte Carlo (MCMC) to this setting. In late 1990 I gave a seminar at the University of Toronto, met assistant professor Michael and learned that he had laid the bases of MCMC for such models in his PhD (at Yale, with John Hartigan).  We immediately started collaborating and generalised the approach and its scope, including hyper-parameter learning, and embarked on putting together papers.  
The initial draft of what became the 1995 JASA paper was defined in the 1990 Duke/ISDS Discussion Paper (\#1990-16) titled {\em Bayesian Prediction and Density Estimation}; we eventually changed the title.  Our 1995 JASA paper should probably have been pushed earlier, but, well, we were both very busy with lots of other things too!  In the couple of years after we met, Michael was settling-in as an junior academic and with his personal life in Toronto.  I was still getting up-to-speed on directing the development of ISDS at Duke while juggling a number of other projects-- including a couple of new collaborations-- and then also welcoming the arrival of our second child, our son Dylan, in 1993.  So, it came out in JASA in 1995 instead of a couple of years earlier as might have been expected. The impact was not, I guess, so seriously impacted by the delay.   In the interim, one of Michael's sole-authored PhD papers came out in JASA 1994, and we had a broader hierarchical Bayes framework with  Peter M\"{u}ller also published in 1994. 
 It has certainly been interesting to see how many applications, extensions and variants have developed over the last 30 years in the so-called BNP (Bayesian nonparametrics) and allied statistics-machine learning literatures.

\begin{figure}[t!]
\centering 
\begin{minipage}{3.2in} 
\centering
\includegraphics[clip,width=3.2in]{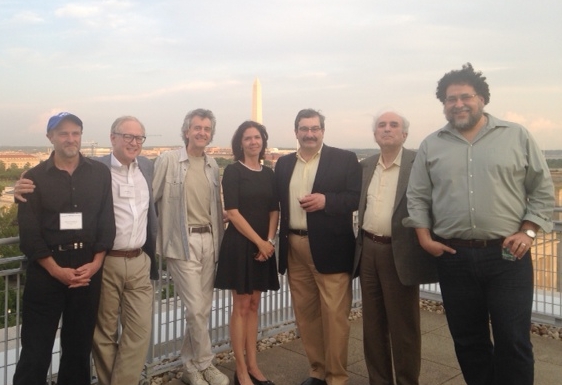}  
\caption*{Mike with friends at the  ISBA–George Box Research Workshop on Frontiers of Statistics (a memorial conference for George Box) in Washington DC, 2014. 
Left to right are Rob McCulloch, Ed George, Mike, Raquel Prado, Refik Soyer, Ehsan Soofi and Hedibert Lopes.
}
\end{minipage}
\hskip0.25in
\begin{minipage}{3.2in} 
\centering
\includegraphics[clip,width=1.5in]{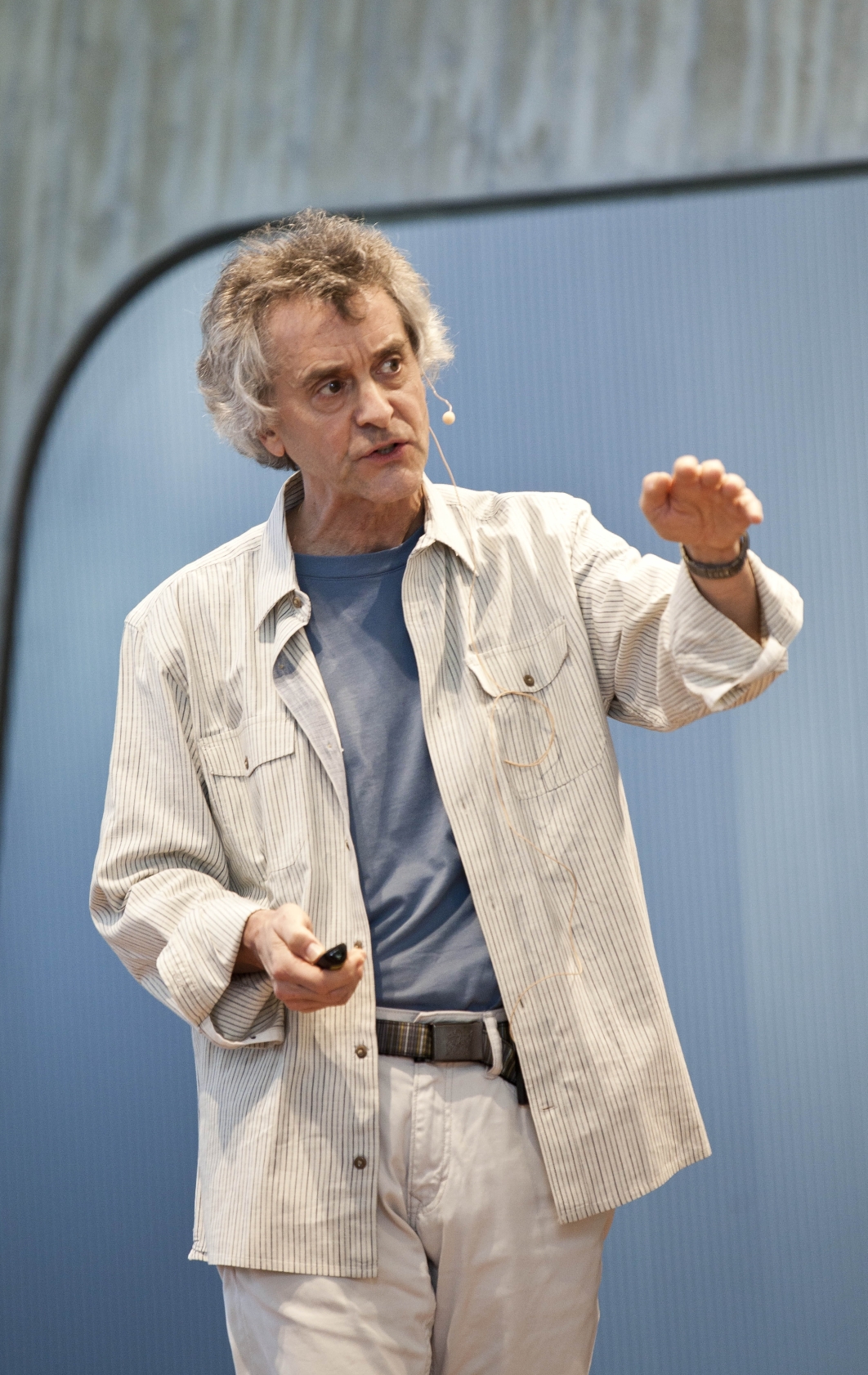}  
\caption*{Mike in full-flow at the BAYESM conference in Vienna, 2014. 
BAYSM (Bayesian Young Statisticians Meeting) is the official conference of j-ISBA, the ISBA section for junior researchers.) 
}
\end{minipage}
\end{figure}

\newpage

\textbf{Hedibert and Filippo}: Recently you worked a lot on the framework of predictive  synthesis. How did you come up with that?

\textbf{Mike}: It is interesting to me that these recent and current areas of research link back so intimately to some of my earlier research from 40 years ago. In the early 2010s, one area of renewed interest for me was in general questions of comparing, relatively calibrating and potentially combining predictive models. This was partly motivated by interests in latent factor and/or graphical models of various kinds, mostly in sequential forecasting and decision contexts, but heavily concerned with model uncertainty. There may be a few or potentially  many models being evaluated in parallel, with all the usual challenges of individual model biases, variations in relative predictive performance across sample spaces, dimensions and time, and of cross-model dependencies (among other challenges including, of course, computation/implementation to begin).  In personal R\&D and linked research with students and others in financial forecasting, and its intersections with the use of similar models in economic applications, I began to explore model combination ideas that-- initially-- went beyond
 traditional Bayesian model averaging (BMA)  to focus on the importance of defining specific predictive goals.  One model may be preferred in terms of BMA when forecasting a collection of macroeconomic series next quarter, but a different model may be more strongly preferred in accurate forecasting of a turning point in one or more series, or the full path of development over future quarters-- the future trajectory-- of a set of series of interest. This focuses attention on specific predictive goals in arbitrating between, and potentially combining, multiple models.
 
In 2013 I ran into a Bank of England (BoE) technical report on \lq\lq generalized density forecast combinations'' 
(later published as Kapetanios {\em et al}, {\em  Journal of Econometrics} 188:150--165, 2015).  This presented a creative  approach 
that defined what we now call outcome-dependent model weighting, with persuasive empirical validation.  The idea is that, for example, one model may be judged as \lq\lq more accurate'' in forecasting national inflation in periods of time when inflation is high, another when inflation is low;  so in \lq\lq weighting models'' predicting future inflation over a period of quarters or years, weights would naturally depend on the as-yet unknown outcomes of inflation over that period.  The BoE group had been able to present a model combination approach that explicitly allows this; and with a convincing example,  I was intrigued. However, the approach was not something I could accept {\em prima facie}; it had no obvious Bayesian rationale, which for me is key to understanding, interpretation, communication, generalisation ... and just to innate interest and stimuli to look at potentially \lq\lq new ideas'' emerging in the literature.   
 So, my main question was whether we could understand the construction as arising from a formal, rational statistical foundation. Around the same time, early discussions with then PhD student Ken McAlinn-- concerning the meaning and scope of the term  \lq\lq model'' in  the traditional Bayesian views of model uncertainty--  led to a connection to  my work on agent (\lq\lq expert'') opinion analysis from the 1980s and early 1990s. That line of research-- that has continued   across areas of statistics, management science and economics-- concerns comparing and combining probabilistic forecasts from multiple \lq\lq agents'', with
origins in earlier  works of Dennis Lindley, Mark Schervish and Christian Genest, among others.
The essential step was then simply matching \lq\lq agents'' with \lq\lq models'' and  building on the foundational (nonparametric and subjective Bayesian) theory of that earlier work. As part of this, it became clear  that the technical constructions used by the BoE group were (almost) justifiable as examples of theoretically implied predictive distributions defined as a formal Bayesian synthesis. That recognition means that we can now understand some of the implicit assumptions behind the approach, and of course place it in a much broader framework where other assumptions-- relevant in different applied contexts-- would generate such a synthesis but with possibly very different technical structures. 

This led to the development of Bayesian predictive synthesis (BPS)-- with basic research including past PhD advisees Ken McAlinn and Matt Johnson.  Revisiting and building on the theoretical foundations, this extended methodology with macroeconomic applications as primary and productive collaboration with past PhD advisee Jouchi Nakajima and econometricians including Knut Are Aasveit.
Some of this has reconciled a number of existing approaches of model synthesis, and  other groups-- across the academic: central banking interfaces, in particular-- have and are exploiting and expanding  the applied methodology with economic and financial applications.

 In core statistical methodology, one theme that evolved is the use of BPS with many models in sequential time series settings, and explicitly addressing defined predictive goals. Some examples in research with those named above and also with past advisees Isaac Lavine and Michael Lindon, again with macroeconomic applications, really highlighted the scope of BPS in admitting \lq\lq goal-oriented'' relative weightings of models. Typically, we build and use models for specific purposes; if specific, articulated forecast goals define the interest, they should be recognised and formally integrated into the analysis. This does mean accepting that different analyses play out with respect to different goals. This view is somewhat contra to the traditional view of adopting one framework, scoring models relatively on some neutral metrics (as BMA does, for example),  and then 
imagining that this is \lq\lq optimal'' for all potential downstream predictions and decisions.

\begin{figure}[t!]
\centering 
\begin{minipage}{3.4in} 
\centering
\includegraphics[clip,width=3.4in]{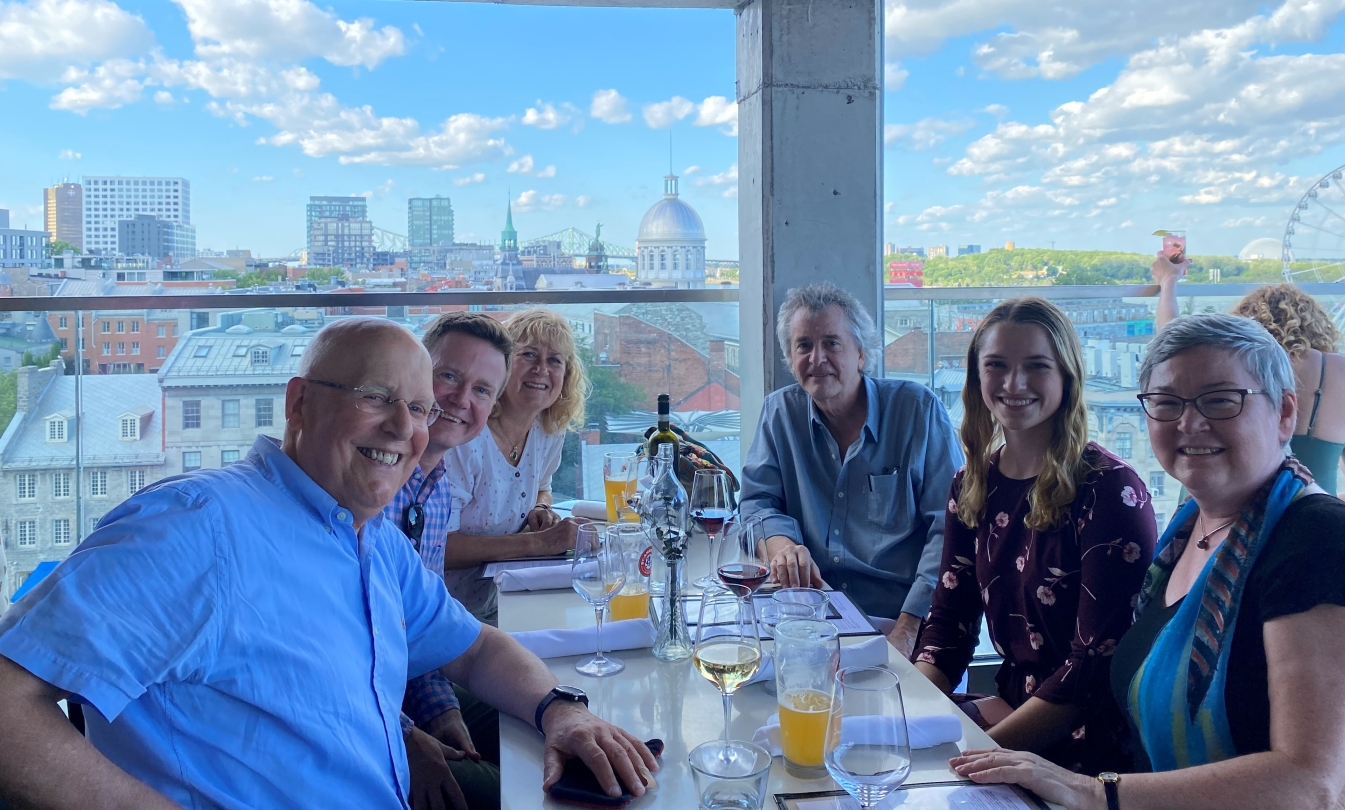}  
\caption*{With friends at the ISBA World Meeting in Montreal, 2022, following Mike's presentation of the  Bruno de Finetti Lecture. Left to right are  Herman van Dijk, James Mitchell, 
Lauren, Mike, Emily Tallman (Duke PhD 2024) and Sylvia Fr\"{u}wirth-Schnatter.
}
\end{minipage}
\end{figure}

\textbf{Hedibert and Filippo}: Your more recent innovations in predictive synthesis involve combining statistical innovations with the classical decision-theoretic approach. Tell us about that.

\textbf{Mike}: 
Adding the  \lq\lq D'' to BPS defined BPDS, i.e., Bayesian predictive {\em decision} synthesis.  Based on my commentary above it is obvious that use of models, and model sets, to inform decisions should factor the decision aspect into the business of marshalling models. Again, we typically build models for purposes, and often the main goals are downstream decisions.  These decisions rely on predictions, so predictive interests are ingredient.  Then,  sometimes it is decision goals and the outcomes of decisions that are of primary interest.   A good example is in financial time series modelling for portfolio decision analysis in investment management and personal decision making.  There is some interest in evaluation and comparison of forecasting approaches based on predictive accuracy in forecasting asset prices and returns, and with defined predictive goals (i.e., which assets, over what time period).  However, it is the roles of derived forecast information in advising investment decisions and the resulting outcomes of those decisions that are primary.  A specific subset of models may (as is typical in this area) generate forecast distributions that seem to differ only modestly in terms of many usual statistical metrics (including BMA). However, the implied optimal portfolio reinvestment decisions-- with respect to what might be, and often are, model-specific utility functions-- can show more meaningful differences.  As a result, the realised outcomes of model-specific informed decisions-- here in terms of a range of  risk and return metrics-- can substantially differentiate models.  I want \lq\lq good'' forecasting models, but my over-riding concern is for models that generate \lq\lq good'' portfolio returns with controlled risk. In other areas, such as in decision making in monetary  policy in central banks, the same general considerations arise but with more of a balance between predictive accuracy and decision outcomes.  There models are sometimes distinguished on foundational, economic bases, and relative predictive performance putatively sheds light on the underlying assumptions across models.  However, the role of decision outcomes in terms of weighting these model-based economic \lq\lq hypotheses'' is ingredient and should be explicit in the statistical analysis, comparison and synthesis of the set of models. 

This line of thinking morphed BPS to BPDS, with initial developments largely involving past PhD advisee Emily Tallman.   The theoretical foundation of BPDS allows relative model weightings to explicitly depend on decision outcomes as well as pure forecast accuracy.  The earlier noted concept of outcome-dependent model weights-- to represent expectations in the future-- also applies.  On this theoretical basis, we are now able to explicitly incorporate expectations of outcomes of current and future decisions into scoring of models; hence the pithy (and hopefully memorable)  labelling of this broader view of Bayesian analysis through BPDS as   allowing for-- and encouraging-- \lq\lq betting on better models''.    Financial portfolio applications with Emily provide persuasive examples of the potential benefits of this broader view of Bayesian model uncertainty in decision contexts.  
Current developments in macroeconomic policy decision settings (with econometricians Tony Chernis and Gary Koop), and related areas of central banking decision making reliant on so-called \lq\lq scenario forecasting'' represent some of what I expect to become central examples of the broader utility of BPDS.

\textbf{Hedibert and Filippo}: You have always worked a lot with private and non-academic organizations as a consultant. How much impact did this have on your academic career?

\textbf{Mike}:  I have enjoyed and professionally thrived on R\&D interactions with a range of \lq\lq private organisations'' (companies, national labs, central banks, various non-profits). While some of this has been rather focused and proscribed consulting, much and most of it has been-- and is-- broader research collaboration enabled via university: industry projects that can and have involved students and postdocs.  My research has substantially benefited through the usual academic channels of public funding through national funding agencies (in the UK and US) and particularly in terms of how these  channels have enhanced-- sometimes enabled-- interdisciplinary collaborations. Coupled with that, my and our interactions in research with industry and commerce have been  substantially important to my own professional development, perspectives on what is interesting in core research as a result of connecting to applications,  and to the evolution of research in areas of my interests and those of our community as a result. This has been, for me, entirely consonant with various themes in my research that have developed through collaborative, interdisciplinary collaborations with academic communities.

Some of my 1980s involvements in corporate and demand/sales forecasting early (groups in ICI and IBM, and market research firms) were particularly formative in terms of my real-world, applied perspectives. Much of that also quickly fed back to motivate core methodological and published research.  Since the early 1990s I have had sustained interactions-- some collaborative research, some advisory-- with investment management groups in banks, hedge funds and the insurance industry.  This again has been stimulating to basic research and involved many students, especially PhD advisees, though internships and then often into careers. Similar comments apply to interactions with biotechnology companies, from the early 2000s and for over a decade of sustained research in genomics.  In more recent years,  some most productive industry-supported research involving numerous PhD (and some MS) students has been with large IT and consumer sales companies, involving ranges of modelling and forecasting challenges. A range of publications over the last decade reflect some of the impact this has had on my own basic thinking about approaches to modelling and computation, on the resulting core
methodology that has arisen and-- through multiple student co-authors-- on the career paths of PhD students. Examples involve  approaches to dynamic network modelling, hierarchical and multi-scale dynamic factor models,  new model classes that morph structure over scales of aggregation, resulting approaches to uncertainty assessment related to rare events,  a number of developments focused on scalability of Bayesian analysis methods exploiting the \lq\lq decouple/recouple'' concept, and-- more recently-- casual prediction in multivariate time series. 

\textbf{Hedibert and Filippo}: Would you then suggest young researchers to keep contacts with the industry?

\textbf{Mike}: I am a strong promoter of academic-industry collaborations for reasons highlighted above. Industry is full of very smart people working on challenging and important topics; certainly the typical goal-oriented foci and shorter time-horizon schedules can be very different than those we define and control as academics. My experiences have generally been positive in identifying  areas where the goals can be aligned, to the benefit of the specific applied collaboration and with feed-back to core statistics research.  Many of my past PhD (and other) advisees moved into industry, sometimes as a specific follow-on from R\&D as a student but always having benefited in maturing as statistical scientists from the industry exposures as students. I have and always will regard this as wholly positive for the individuals, their departments and the profession.

\section{Looking ahead}

\textbf{Hedibert and Filippo}: It appears that you will attend quite a few conferences  in 2025-- so Emeritus status has not slowed you down! How do you see the next few years?

\textbf{Mike}: At time of writing in August 2025, the count is 6 conferences (or workshops and research retreats) in terms of travelling for \lq\lq in-person'' participation, and then 2 or 3 others where I appear and participate virtually.  Conference engagement has always been a main part of my professional calendar-- critical for connecting and reconnecting with colleagues, students, alumni and friends, and making new connections.  I do find it hard to turn-down speaking invitations, especially at smaller and 
focused research workshops, with the opportunity to communicate and discuss some of what I find interesting at the time.  Over the years, a number of themes in my research have been either sparked or redirected based on Q\&A at conference presentations. A  number of collaborations have been so initiated.  Supporting presenting students and other collaborators has always been a priority motivating me to attend.   Then, having been Emeritus for  almost a year, I can say that I think I am making the most of the reduction in time and energy commitments that a full-time faculty position entails. I believe I have seen this in terms of research advances this year so far-- especially based on time to just sit-back and think a little more now and again.  I have a number of \lq\lq retirement'' hobbies, several of which just happen to involve statistics R\&D!   Complete freedom to travel almost anytime, anywhere and at-will is an additional positive, and of course that links to conferences.   An important point is that the logistics of R\&D collaborations-- including continuing advisory roles with PhD students-- have been significantly impacted by remote/virtual meeting technology.  It is hard to call the virtual revolution a silver-lining of the Covid19 pandemic, but over the last few years we have exploited this in very positive ways that were, perhaps, earlier somewhat unforeseen. The reality is that we can be almost anywhere, in any time zone, and still maintain routine and 
productive interactions.

 \begin{figure}[t!]
\centering 
\begin{minipage}{3.2in} 
\centering
\includegraphics[clip,width=1.8in]{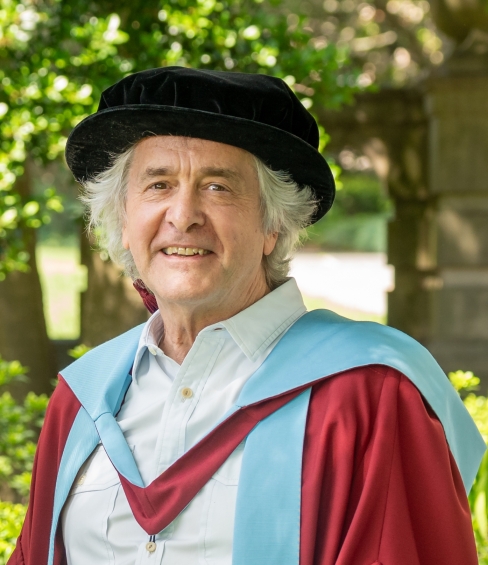}  
\caption*{Mike at Duke University graduation, May 2024.
}
\end{minipage}
\end{figure}

I don't see any reason to expect any real departure from this trajectory in the near-term.  Lauren
 and I have a very busy travel schedule for the coming several months, and an already  \lq\lq pencilled-in'' series of dates and locations associated with a number regular and one-off conferences and workshops over the next couple of years.
 So long as the invitations keep coming, with the expectation of engaging with existing colleagues and friends and encountering new ones, and so long as we are able to book online and carry passports, we'll be travelling.

\textbf{Hedibert and Filippo}: Which questions are you currently interested in? Do you have any major books or projects under way?  

\textbf{Mike}: Apart from politics and international affairs,  some questions of current interest link to a few related projects-- among my hobbies. No book projects on the horizon, and none expected!    The only potentially relevant considerations are my teaching materials built and evolved over the years, including several hundred pages of detailed \lq\lq book'' material (10 point font, single spaced) with hundreds of exercises on introductory graduate-level 
probability and statistical modelling, and on 
decision analysis and decision theory, with a broad statistical and applied purview. 
Along with code and data for most of the exercises and more, these might  naturally have taken formal book form  $15{-}20$ years ago.  Nowadays, this could define a nice project for someone to reformat and curate it all, and make up public web sites; but, while I would be very pleased to see it happen, it is not something I can now imagine finding the time to redevelop in traditional book form, at least near-term. 

On current research interests, I will mention a few  current themes and projects.  First, I maintain an interest and research activity in multivariate forecasting and dynamic models, especially in the further development and application of flexible, scalable simultaneous graphical dynamic models (SGDLMs). One current theme here is for financial time series and volatility forecasting, involving new theory for multi-scale analyses that current PhD advisee Patrick Woitschig has recently introduced, and that we are now integrating with SGDLMs.    The applied setting involves portfolio decisions, so there is ingredient interest in new, practically motivated utility functions for more-than-stylised personal and corporate decision analysis ({\em hint}: always avoid unbounded utility functions, and always anchor decisions in baselines/benchmarks). Some of the perspective driving that line of research comes from collaboration with financial practitioners. 
A second related area is the development of SGDLMs as a context for advancing counterfactual prediction for putatively causal forecasting in multivariate time series. In addition to generating new, flexible methodology towards that main goal, this has expanded to explicate new theory relating sparse (and dynamic) graphical models and sparse (dynamic) factor models through SGDLMs; innovations of PhD advisee Luke Vrotsos are driving this.   Some of the counterfactual/causal research here has extended some recent papers with current and past PhD students Kevin Li and Graham Tierney (in part related to early industry:academic collaboration in marketing in challenging settings of consumer demand/sales response to promotions).  The new SGDLM-based approach really opens up much more flexible modelling framework, and also nicely ties into potential future interests in applications in economic policy settings with links to so-called scenario analysis.

A second main area is advancing BPDS.   Here one main focus is in formal Bayesian forecasting and decision analysis in macroeconomics, especially related to monetary policy and in collaborations with central banking researcher Tony Chernis, econometrician Gary Koop and past PhD advisee Emily Tallman.  Modelling for economic forecasting that aims to advise policy decisions in such contexts involves putative decision variables, such as the interest rate targets that national central banks define for cost of money to commercial banks, and that massively impact financial markets and economic developments. One catch is that realised future paths of such variables result from the complex economic-financial system, so decision variables are also outcome variables.  One part of this research has defined clarity on this core issue and the impact on conditional forecasting -- a.k.a. \lq\lq what-if?'' or \lq\lq scenario'' forecasting-- in this area.  Then, the major focus of this program is the development of BPDS to manage and analyse forecast information and associated optimal policy recommendations from multiple, potentially competing though also often highly-related models.  While BPDS has established credentials in financial portfolio analysis, that is a setting in which models are used to forecast and then decision analysis applies to the synthesis.  In contrast, policy intervention analysis is a setting of sequential statistical optimal design, or so-called reinforcement learning where predictions from each of a set of models are conditioned on ranges of values of the putative control/decision variable. This requires new thinking about utility function choices and 
generates  novel analytic and computational challenges. The questions of managing, understanding and synthesising information from multiple sources and models-- and the paths to then exploring optimal decision paths for policy makers-- are increasingly topical across central banking research groups worldwide as they are in other areas.

\textbf{Hedibert and Filippo}: Do you also have ongoing research projects in collaborations outside academia?

\textbf{Mike}: A parallel theme in my current research concerns related questions of scenario forecasting that have recently become quite high-profile, partly as a result of the report of former US Federal Reserve chairman Ben Bernanke on the Bank of England's policy-informing forecasting enterprises
(April 2024, BoE, Forecasting for monetary policy-making and communication at the Bank of England: A review). This is currently a main theme 
in discussions within and between research groups in central banks-- including the BoE, the Federal Reserve Board (FRB), the European Central Bank (ECB) and the International Monetary Fund (IMF). 
Some of the core questions here have to do with the assessment and integration of multiple predictive perspectives: scenario forecasting has been associated with forecasts (often fairly sparse summaries of probabilistic forecast distributions) having economic bases in assumed future economic developments. That is, scenarios are hypotheses with underlying economic \lq\lq theory'' generating \lq\lq stories'' that can be communicated, discussed, argued about and relatively assessed on foundational bases. On the other hand, more empirical econometric (i.e.,  statistical) models define forecast distributions  that are often more customised to available data and, in particular, can be more responsive to unpredicted changes and natural time-variation in relationships among economic time series.   In collaboration with Tobias Adrian (IMF), Domenico Giannoni (IMF and Johns Hopkins University) and Matteo Luciani (FRB), we have been and are developing a formal statistical framework for bridging scenario story-based forecasting and more empirical and robust statistical model-based forecasting.  This current and topical context in the policy domain has motivated new concepts and theory-- related to BPDS but with somewhat different goals and resulting methodology.  I imagine that this will be a growing and active research area for me-- with an expanding number of collaborators-- for some time to come.
 
 \textbf{Hedibert and Filippo}: What are the connections of your ongoing work with fields that are now rapidly evolving? We are thinking about nonparametric Bayes, generalized Bayesian methods, predictive contractions, etc.

\textbf{Mike}: On the BPDS themes, apart from the broader view of Bayesian analysis it represents, the theoretical foundations are wholly in the nonparametric domain. 
Bayesian nonparametrics (BNP) has been a very major areas of expansion-- across statistics and machine learning rather broadly-- over the last 25 years and more. However, much of the expansion has been in probabilistic models with infinite-dimensional parameters. While the field is irretrievably referred to as \lq\lq nonparametrics'',  the more traditional use of the term is in settings involving only partial-- often very sparse-- assumptions about \lq\lq data generating mechanisms'' and resulting classes of statistical models explored.   This is true in frequentist statistics as it was in the early years of development among Bayesians.  The agent opinion analysis genesis of BPDS is firmly based in, and aligned with, this more traditional use of the term.  As modern BNP continues to grow and develop, I would encourage new researchers, in particular, to consider a step-back from the fascinating activity in increasing large models, increasing complexity of model structures with infinite-dimensional parameters-- and the consequent challenges to understanding what assumptions really matter and might drive \lq\lq results'' without such understanding.  BPDS is just one setting that is more anchored in the traditional perspective of making limited assumptions-- understanding and defending them-- and one that is open to new ideas and investigation, theoretical and methodological.

Our conversation above directly touches on the question about the recent surge of research focused on prediction 
(including martingale posteriors, generalised and so-called \lq\lq robust'' Bayes).  I regard these as interesting and important areas of development, for a number of reasons. First, some of the developments in \lq\lq focused prediction'' that intersect this general area are explicitly linked to BPS, as defined in some of our first developments in time series forecasting with many models.  That includes the developments of mixture-model weightings with specific forecast goals defining utility-- or score-- functions to apply to each model (with past PhD students Matt Johnson, Isaac Lavine and Michael Lindon), and to multivariate dynamic latent factor models within which individual models are represented as defining latent processes (with past students Ken McAlinn, Jouchi Nakajima and others).  As noted earlier in this conversation, the BPS perspective and theoretical framework encompasses some of the recent developments in the literature on the use of utilities to define synthetic \lq\lq likelihood functions''.  The  theory of BPS also admits replacing \lq\lq models'' with \lq\lq parameters'', and so applies to problems more consonant with the focus of some of this recent literature on parameter inference, with a core aim of that literature being to avoid specifying full sampling models.   This opens up opportunities for next-steps  explicitly recognising that BPS theoretically underpins some of the recent developments. More importantly, then, is the integration with decision analysis that the more general BPDS framework represents. Inference about a parameter can be of interest -- so long as it is explicitly defined, interpretable and grounded in an applied context.  However, much and most of applied statistics is about prediction and decisions. As a predictivist, I am emphatically supportive of-- and centrally engaged and interested in-- some of the noted developments that emphasise prediction. But! Don't forget the decisions ... the \lq\lq Yang'' to the predictive \lq\lq Yin'' of statistical analysis. 

\textbf{Hedibert and Filippo}: How much have you seen the academic world change over the years? And what do you expect for the near future?

\textbf{Mike}: This partly relates to the nature of academic statistics linked specifically to the previous comments on industry:academia interactions. 
The last few decades have seen massive change in our profession generally, and particularly in terms of the centrality of statistics-- broadly understood to intersect with data science and machine learning--  across many industries and other non-academic organisations.   University statistics programs must evolve, and have evolved and adapted. This has  partly come through processes of natural evolution of what and how we teach: computational technology, aspects of computer science and \lq\lq big data'' perspectives permeate our research and very naturally our teaching. Part of the change in many departments has been more strategic-- the expansion of Master's programs in statistics and data science is the key example.  At Duke we have a large and vibrant (and challenging) MS program whose graduates are predominantly oriented towards industry, though with a substantial-- and laudably increasing-- number moving to PhD programs too.  Across many department this has naturally impacted on faculty time and intellectual energy commitments,   but is in my view a critical response to the increasing demand for statistical thinkers and doers that is likely to continue for years. If core statistics departments and their research faculty are not engaged in responding to the market and societal demands, others will-- and this may run us into the challenges of an increasing \lq\lq statistics'' workforce that has less deep and detailed experience in core statistical thinking than perhaps we would regard as relevant and desirable.

Other aspects of these changes are more challenging and to a degree concern me. One challenge is that, as new and more technological topics grow in teaching and research programs, they jockey for space with more traditional and foundational topics. All PhD students should have opportunities to take courses-- or mentored  independent studies-- in core areas of decision analysis and aspects of decision theory,  and aspects of sampling and design, for example, along with mainstream modelling, inference and data analysis.  And, of course, time series!  However, statistical science is now so very broad  that any program must be selective, and continue to customize around core faculty interests and expertise.  There is, of course,  much to benefit from  increasedly diverse perspectives. I do, however,  have concerns that new faculty with core backgrounds in foundations and mainstream statistics are increasingly under-represented  due to the expansion of algorithmically focused, and technologically driven, machine learning perspectives at PhD level. 

Many statistics PhD graduates  move onto interesting and rewarding 
non-academic careers, diluting the flow of top thinkers to academic positions. 
How to address this need to substantially grow the numbers of PhDs moving into university faculty positions in statistics and allied disciplines? 
 Some might take the view that we need to more aggressively promote the traditional path to faculty positions (perhaps following a postdoc) and downplay, to a degree, potential engagement of PhD students in industry internships and collaborations.  This is simply orthogonal to my view, and just wrong.   First and foremost, our job as advisors is to advise, and to help each student progress in whatever directions they are motivated to explore; opening doors to industry collaborations and internships is a big part of that.  The only rational response is to very substantially increase the numbers of PhD students in core statistics programs worldwide.

\textbf{Hedibert and Filippo}: How do you place the connection between AI and Statistics? Do you see opportunities or threats?

\textbf{Mike}: I have never been overly concerned about \lq\lq threats'' in thinking about how innovations in technology and research in other areas might or might not impact on our discipline. I like to reflect on the potential opportunities such developments present.   The last few years of developments in AI in the public domain, especially related to consumer-targeted LLM-based tools, is huge and presages mega steps in the current phase of the new industrial revolution.  I am not currently a major consumer, but do dabble a bit and will increasingly so. It is easy and quick to generate potential leads on queries of interest; the challenge currently, for users, is to winnow the signal from the noise in the results.  On the more specific context of machine learning for prediction (deep learning etc.) which is where perhaps more immediate  questions about relationships with statistical analysis arise.   This has been a terrific few years of massive advances in scalable and hugely flexible, really nonparametric models and their customisation to contexts.  With very large and rich training data, improved accuracy of prediction on the \lq\lq next'' set of samples from similar contexts is undeniably anticipated.  Then, some central statistical questions arise. What about uncertainty characterisation?  What about generalisation and transfer to \lq\lq new'' contexts that are really not new, but random perturbations of contexts represented in the training data; however large and rich the latter may be, random effects matter! 

Much of my own work over the years has been concerned with sequential learning and forecasting, typically with specific decision goals in mind.  From my early years in developments of sequential monitoring of classes of models (that could be regarded as, or replaced by, \lq\lq algorithms'')  I have always been concerned about encountering interesting events, \lq\lq weird'' data, news and information that seems relevant to the forecast and decisions addressed down the road.   The philosophy is that of explicit recognition and integration of the modellers and decision makers into the process of learning and forecasting, with routine monitoring and openness to potential interventions to modify and adjust models and hence their forecasts in face of new information The latter must recognize the potential to have to respond to  \lq\lq rare events''. This perspective has underpinned methodology development and many applications (including my own personal forecast enterprises and decision making as well as that in collaborative research). 

 At the start of 2022 my involvement in collaborative R\&D with two major consumer companies and organisations expanded, and of course the  very major concerns had to do with the reality that the Covid19 pandemic hit. This was a unique series of events in the  fabric of society worldwide and, within that, in the 
history of modern economies and markets: a Black Swan event. Models and algorithms-- however \lq\lq intelligent''-- that were wholly based on historical training data and pattern matching had limited, if any, validity in terms of informing the responses of companies, governments and other agencies in predictions and resulting decisions over coming months.  The impact on national and multinational companies was profound. Short-term forecasting of consumer demand, from the very micro (one pair of this particular brand, size, style of jeans, or one 2lb bag of this particular brand and roast of coffee beans) was upended by simply unique dislocation of the labour forces, and of the international, national and regional supply chains.   
Overlaying this were substantial, but really unique and unpredictable changes in consumer behaviour. 
The impact on forecasting at intersecting aggregate levels-- \lq\lq regional'' in terms of geography,  consumer market sectors, and in time-- was enormous in terms of cascading uncertainties.  The feed-through to revenue forecasting  and high-level corporate planning and decisions were profound. 
 There were many months of unique behaviour in markets and local economics as a result of the disruptions, and otherwise proven and apparently rational automated systems for digesting data and returning predictions based on recent/past data tended to fail.  This recent lived experience exemplifies the critical needs for model systems to be open to user control and intervention, to adequately adapt to volatility in environments that move out of \lq\lq regions of experience'', to allow and encourage informed integration of new and emerging data and  (user, personal, subjective) information from multiple sources, to formally address appropriate characterisation of uncertainties, and especially to admit the potential for \lq\lq rare events'' as being profoundly important. This contrasts what I regard as some of the core elements of statistical thinking and analysis, quite generally, with more automated ML-style and algorithmic approaches to prediction.    While the technological advances in ML/AI in recent years are quite profound and will surely be central to much of what we do and think is useful in coming times,  the roles and relevance of core statistical thinking-- as I have tried to articulate, from my personal, subjective Bayesian perspective here-- are, well foundational and immutable.  As we move ahead, I hope and expect to see more aggressive and overt development of foundational  \lq\lq uncertainty management'' perspectives in evolving AI systems, to the benefit of future users and to society broadly. 

I will add that I have for years been very fond of completely re-wording Dennis Lindley's invocation of what he named \lq\lq Cromwell's Rule'';   {\em Always be prepared to be surprised!}


\section*{References and More}
\noindent Mike's publications (1981--present) are listed and linked at \href{https://www.stat.duke.edu/~mwest/}{stat.duke.edu/$\tilde{\,}$mwest}. Also there are links to past advisees and associates, and more.

\end{document}